\documentclass[prd,aps,floatfix,superscriptaddress,twocolumn,nofootinbib]{revtex4-1}

\usepackage{amssymb,amsmath,latexsym,booktabs}
\usepackage{graphicx,float}
\usepackage{bbm}
\usepackage[usenames,dvipsnames]{xcolor}
\usepackage{tikz}
\usetikzlibrary{arrows,decorations.pathmorphing,positioning}

\newcommand{\bea}{\begin{eqnarray}}
\newcommand{\eea}{\end{eqnarray}}
\definecolor{olivegreen}{rgb}{0,0.6,0}

\usepackage[english]{babel}
\usepackage{microtype}
\usepackage{hyperref}

\begin{document}   

\title{Semi-supervised deep learning for postmerger binary neutron star waveforms}
\title{Using machine learning to parametrize postmerger signals from binary neutron stars}

\date{\today}
\author{Tim Whittaker}
\affiliation{Department of Physics \& Astronomy, University of Waterloo, Waterloo, ON N2L 3G1, Canada}
\affiliation{Perimeter Institute for Theoretical Physics, Waterloo, Ontario N2L 2Y5, Canada}
\author{William E.\ East}
\affiliation{Perimeter Institute for Theoretical Physics, Waterloo, Ontario N2L 2Y5, Canada}
\author{Stephen R.\ Green}
\affiliation{Max Planck Institute for Gravitational Physics (Albert Einstein Institute) Am M\"{u}hlenberg 1, 14476 Potsdam, Germany}
\author{Luis Lehner}
\affiliation{Perimeter Institute for Theoretical Physics, Waterloo, Ontario N2L 2Y5, Canada}
\author{Huan Yang}
\affiliation{University of Guelph, Guelph, Ontario N1G 2W1, Canada}
\affiliation{Perimeter Institute for Theoretical Physics, Waterloo, Ontario N2L 2Y5, Canada}

\begin{abstract}
  There is growing interest in the detection and characterization of
  gravitational waves from postmerger oscillations of binary neutron
  stars. These signals contain information about the nature of the
  remnant and the high-density and out-of-equilibrium physics of the
  postmerger processes, which would complement any electromagnetic
  signal. However, the construction of binary neutron star postmerger
  waveforms is much more complicated than for binary black holes: (i)
  there are theoretical uncertainties in the neutron-star equation of
  state and other aspects of the high-density physics, (ii) numerical
  simulations are expensive and  available ones only cover a small fraction of the
  parameter space with limited numerical accuracy, and (iii) it is
  unclear how to parametrize the theoretical uncertainties and
  interpolate across parameter space. In this work, we describe the
  use of a machine-learning method called a conditional variational
  autoencoder (CVAE) to construct postmerger models for hyper/massive
  neutron star remnant signals based on numerical-relativity
  simulations. The CVAE provides a probabilistic model, which encodes
  uncertainties in the training data within a set of latent
  parameters. We estimate that training such a model will ultimately
  require $\sim 10^4$ waveforms. However, using synthetic
  training waveforms as a proof-of-principle, we show that the CVAE
  can be used as an accurate generative model and that it encodes the
  equation of state in a useful latent representation.
\end{abstract}

\maketitle 

\section{Introduction}

With 90 detections of gravitational waves from merging compact
binaries since
2015~\cite{LIGOScientific:2018mvr,LIGOScientific:2020ibl,LIGOScientific:2021djp}, including
up to five involving neutron
stars~\cite{LIGOScientific:2017vwq,LIGOScientific:2020aai,LIGOScientific:2020zkf,LIGOScientific:2021qlt},
a key goal of gravitational-wave astronomy is to maximize the
scientific pay off of current and future detections. Neutron star
mergers are especially interesting since they have the potential to
elucidate many physical phenomena, from gravitational to nuclear
physics.

There have been several proposals to combine observations from
multiple events to increase the signal-to-noise ratio (SNR) of
specific types of information
(e.g.,~\cite{Meidam:2014jpa,Yang:2017xlf,Yang:2017zxs,Brito:2018rfr}).
For instance, by analyzing different stages of a non-vacuum binary
merger, one can extract the signatures of tidal effects, the rate of
relaxation to equilibrium of different multipolar perturbations, the
connection of a binary's intrinsic parameters to the object formed
after the merger, and potential new measurements of the Hubble
constant using only gravitational waves (e.g.~\cite{Messenger:2011gi,
  bustillo2020higherorder, Miao:2017qot}).  Such efforts would help
reveal the equation of state (EOS) of neutron stars, probe possible
deviations from General Relativity, allow for stringent tests of the
{\em final state conjecture}\footnote{Sometimes also referred to as
  the ``generalized Israel conjecture.''}~\cite{1969NCimR...1..252P},
explore the behavior of hot matter at supra-nuclear densities, and
help elucidate the inner engine of gamma ray
bursts~\cite{Sekiguchi_2011,Radice_2017,Most_2019,Paschalidis_2017,Ruiz_2016,Dall_Osso_2014}.

In this work, we focus on the post-merger stage of binary neutron star
(BNS) coalescence.  In particular, we seek a method to extract key
properties of the hyper/massive neutron star that results if a prompt
collapse to a black hole is avoided.  This occurs if the total mass
$M$ of the system is not too high with respect to the maximum allowed
mass $M_{\rm{max}}$ of a nonrotating star with the same EOS
[$M_T\simeq (1.2-1.7) M_{\rm{max}}$], see
e.g., Refs.~\cite{Shibata_2019,Hotokezaka_2011,Bauswein_2013}. The
gravitational radiation from this stage, sourced by the oscillating,
differentially-rotating object produced by the merger, is rich in
information about the hot EOS of the system, which can be exploited,
in particular, to measure the Hubble constant, guide the understanding
of the central engine of intense electromagnetic outbursts, and
determine the maximum mass of neutron stars.

Achieving this goal will require both knowledge of the waveform
produced during this stage, as well as improved sensitivity of
detectors at the high frequencies (1--4 kHz) that characterize such
waveforms~\cite{Miao:2017qot,Martynov:2019gvu}.  A theoretical
understanding of the expected waveform not only aids in detection, but
is crucial for the physical interpretation of such signals. Naturally,
the degree of completeness of such knowledge, together with the
sensitivity of the detector, impacts the depth of the analysis that
can be carried out. For instance, advanced LIGO \cite{ALIGO} and advanced Virgo \cite{VIRGO:2014yos} reported no such signal
following GW170817~\cite{LIGOScientific:2017fdd}, consistent with
theoretical expectations that it would be have been too weak to
detect. Indeed, the study~\cite{LIGOScientific:2017fdd} searched for
extra power in spectrograms, but even with detailed models a
sensitivity three times greater than aLIGO design would be required
for the detection of the leading mode in the aftermerger
gravitational-wave signal~\cite{Torres-Rivas:2018svp}.

Building models for BNS postmerger signals is challenging for several
reasons. First, a BNS merger is a highly nonlinear event, which can only
be treated numerically. BNS merger simulations are expensive compared to
those of binary black holes, since in addition to solving the Einstein
equation, they must incorporate magnetohydrodynamics and microphysics. 
Typically, this means that BNS simulations converge at lower
order in discretization length compared to vacuum simulations. The BNS
merger also gives rise to small-scale features, such as turbulence, which
are difficult to resolve numerically, but which can significantly affect
the gravitational-wave phase and ultimate fate of the remnant. It is
typical for BNS simulations to develop order one phase errors in the
gravitational wave signal soon after the stars come into contact.
Finally, simulations must be sufficiently long and accurate to smoothly
match to perturbation theory in the early inspiral as well as capture the
after-merger behavior. Consequently, the number of numerical simulations
available at present from which to build a model is limited, and these
simulations involve uncertainties not present in binary black hole
simulations.

Any model must cover the full parameter space for BNS systems, which
in principle includes all parameters for binary black holes, plus --in particular-- the
EOS, which for a cold neutron star gives the pressure as a
function of energy density. During the inspiral the EOS
manifests through the tidal deformabilities of the individual neutron
stars, which can be measured through their impact on the phase
evolution~\cite{Flanagan:2007ix}. Postmerger gravitational waves probe
a hot, dense, high-mass regime complementary to the inspiral. 
Using numerical simulations, the dominant postmerger frequency
$f_{\text{peak}}$ has been connected to the radius
$R_{1.6}$ of a nonrotating star of mass
1.6~$\mathrm{M}_\odot$~\cite{Bauswein:2012ya}. However this
relation is not exact, but rather depends weakly on the average mass
of the neutron stars~\cite{Rezzolla_2016} and mass
ratio~\cite{Lehner:2016lxy}.  The signal is further complicated by the
presence of secondary modes, which could contain additional EOS
information. Thus, in addition to the challenges of limited numbers of
simulations, and uncertainties inherent in simulations, a third
challenge is to parametrize the EOS in a manner that would facilitate
inference.

In this work, we address these challenges using a deep-learning
technique called a conditional variational autoencoder
(CVAE)~\cite{kingma2013autoencoding} to build a distributional
latent-variable model for postmerger signals $h$. The basic idea is to
partition the parameters characterizing the signal into two sets. The
first set consists of those parameters $\theta$ that we have direct
access to from simulations and that we know \emph{a priori} should
form part of the characterization of the system. For a postmerger
signal, $\theta$ could include, e.g., the total mass and spin of the
system. The second set of parameters---the latent variables
$z$---includes the EOS as well as any other physics-modeling or
numerical differences between simulations. It is \emph{a priori}
unclear how best to parametrize these properties of the system and
simulations, so this task is left to the CVAE. During training we do
not provide any information about latent parameters, rather the CVAE
learns to use $z$ to efficiently represent differences in training
waveforms $h^{(i)}$ that are not accounted for in the parameters
$\theta^{(i)}$.

Through training, the CVAE learns a model $p(h|z,\theta)$ for the
waveform $h$ given $z$ and $\theta$. The CVAE also learns an
``encoder'' model $q(z|h,\theta)$ for the latent variables $z$,
conditioned on $h$ and $\theta$. Using the encoder to identify latent
variables with numerical simulations, we show that $z$ encodes
information about the EOS in a useful way. Given a signal, one could
then perform inference jointly over $\theta$ and $z$ using
$p(h|z,\theta)$ to determine the EOS and distinguish modeling
uncertainties in simulations.

Existing postmerger models have been designed mostly to facilitate the
extraction of ringdown spectral information, guided by the empirical
$R_{1.6}$--$f_{\text{peak}}$ relation. Past works have included the use of
principal component analysis~\cite{Clark:2015zxa}, phenomenological waveform
modeling~\cite{Bose_2018}, agnostic modeling such as
BayesWave~\cite{Cornish_2015}, analytic time domain waveforms informed by
numerical simulations~\cite{Breschi:2019srl}, and searches for specific
principal frequencies tied to the physical parameters
(e.g.~\cite{Bauswein:2015yca,Hanauske:2016gia,Lehner:2016lxy,Lehner:2016wjg,Easter:2018pqy,Bauswein:2019ybt,Bernuzzi:2020tgt,Soultanis:2021oia,Tsang:2019esi,Takami:2014tva}.)
A major advantage of the CVAE is that it learns automatically to connect
features of the waveform to aspects of the EOS, rather than depending on
fortuitous discoveries of empirical relations. In this way, it can be applied
more generally, and, with sufficient training data, has the potential to
discover new relations between the EOS and the waveform.

There have been several recent applications of machine learning to
waveform modeling~\cite{Chua:2018woh,Khan:2020fso,Chua:2020stf} (see
also~\cite{Cuoco:2020ogp}). These approaches use neural networks to
interpolate a set of training waveforms across parameter space. Since
waveform generation requires forward neural-network passes, these
models are fast, and they furthermore allow for rapid computation of
derivatives with respect to model parameters, facilitating
derivative-based inference algorithms. Our CVAE-based method
generalizes these approaches to include a latent space for
representing properties of the training waveforms for which a
convenient parametrization may not be known in advance. We also note
that CVAEs have in the past been applied to gravitational
waves~\cite{Gabbard:2019rde}, but to the problem of parameter
estimation rather than waveform modeling.

This paper is organized as follows. In Sec.~\ref{sec:framework}, we 
outline our basic approach and introduce the CVAE. Then in Sec.~\ref{sec:imp}, 
we build a simplified CVAE model $p(h|z,M)$, depending only on the
total mass and the latent variables, for postmerger waveforms using
numerical data from simulations. We find that the number of available
simulations is at present too small to successfully train the model, so
in Sec.~\ref{sec:extended}, we fit instead to artificial waveforms. 
By examining the latent space, we find evidence that the neutron star compactness is encoded in the latent space. We discuss the results of this study
in Sec.~\ref{sec:discuss}, and
estimate that in practice 10,000 simulations\footnote{We assume here the simulations are of similar quality and in the convergent regime during the postmerger.} will be needed to begin to
successfully train a CVAE postmerger model which resembles the numerical waveforms we currently have.

\section{CVAE framework}
\label{sec:framework}
Suppose we have a set of pairs $(\theta^{(i)}, h^{(i)})$ arising from
BNS simulations, where $h$ denotes the signal waveform, and $\theta$
denotes the binary system parameters, in particular the constituent
masses (and potentially the spins, although we will not consider them
here).  However, the simulations have additional underlying parameters
not included in $\theta$, including the EOS, discretization errors,
and additional modeling choices for the matter physics, which would
confound an attempt to build a distributional model $p(h|\theta)$. To
capture the dependence on these additional parameters, we suppose that
they can be represented by a set of latent variables $z$, and we
augment our model by conditioning on these as well, i.e.,
$p(h|\theta,z)$.

We treat the latent variables as \emph{a priori} unknown, and we do not provide them when building the model. Rather, our aim is to learn
a useful latent representation (of the EOS, etc.)~based on patterns in
the training waveforms. To do so, we fix a suitably restrictive
parameterized form for $p(h|\theta,z)$, typically a Gaussian
distribution, with mean and covariance specified as outputs of a
neural network (with input $( \theta,z)$). We also fix a prior $p(z)$,
typically standard normal. Only by encoding information in $z$ can the
marginalized distribution, now a Gaussian mixture,
\begin{equation}\label{eq:marginalized}
  p(h|\theta) = \int \mathrm{d}z \, p(h|\theta,z) p(z),
\end{equation}
be sufficiently general to represent the training data.

Given parameterized forms for $p(h|\theta,z)$ and $p(z)$, we would
like to tune the neural-network parameters to maximize the likelihood
that the training outputs $\{h^{(i)}\}_{i=1}^N$ came from the inputs
$\{\theta^{(i)}\}_{i=1}^N$ under the model $p(h|\theta)$, i.e., we
would like to minimize the \emph{loss} function,
\begin{equation}
  L_{\text{MLL}} \equiv \mathbb{E}_{p_{\text{data}}(\theta)} \mathbb{E}_{p_{\text{data}}(h|\theta)}\left[-\log p(h|\theta)\right].
\end{equation}
We use $\mathbb{E}_p$ to denote the expected value over some distribution $p$. However, the integral over $z$ in Eq.~\eqref{eq:marginalized} is
intractable, so this cannot be evaluated.

To obtain a tractable loss, we use the variational autoencoder
framework~\cite{kingma2013autoencoding}. Intuitively, the integral in
Eq.~\eqref{eq:marginalized} could be made tractable if we knew which
$z$ contributes most for given $h$ and $\theta$, i.e., if we had
access to $p(z|h,\theta)$. However, this density is also
intractable. As an approximation, one therefore introduces an
\emph{encoder} distribution $q(z|h,\theta)$. Then
\begin{align}
  \log p(h|\theta) ={}& \mathbb{E}_{q(z|h,\theta)} \log p(h|\theta) \nonumber \\
  ={}& \mathbb{E}_{q(z|h,\theta)} \log p(h|z,\theta) - D_{\text{KL}}\left( q(z|h,\theta)\| p(z) \right)\nonumber \\
                      & + D_{\text{KL}}\left( q(z|h,\theta) \| p(z|h,\theta) \right).
\end{align}
This expression involves the Kullback-Leibler (KL) divergence,
\begin{equation}
  D_{\text{KL}}(q(x)\|p(x)) \equiv \mathbb{E}_{q(x)} \log \frac{q(x)}{p(x)},
\end{equation}
which is nonnegative and vanishes if $q = p$. We therefore take the
loss function to be
\begin{align}\label{eq:L}
  L &\equiv \mathbb{E}_{p_{\text{data}}(h,\theta)} \left[ - \mathbb{E}_{q(z|h,\theta)} \log p(h|z,\theta) \right. \nonumber\\
  &\qquad \qquad \qquad \left. + D_{\text{KL}}\left( q(z|h,\theta)\| p(z) \right) \right] \\
  &= L_{\text{MLL}} + \mathbb{E}_{p_{\text{data}}(h,\theta)} D_{\text{KL}}\left( q(z|h,\theta) \| p(z|h,\theta) \right). \label{eq:Lcompare}
\end{align}
The expression in Eq.~\eqref{eq:L} is now tractable since we avoid evaluating the integral in Eq.~\eqref{eq:marginalized}. By Eq.~\eqref{eq:Lcompare},
if $q(z|h,\theta)$ is identical to the posterior $p(z|h,\theta)$ then
the CVAE loss function is equal to the maximum-log-likelihood loss.

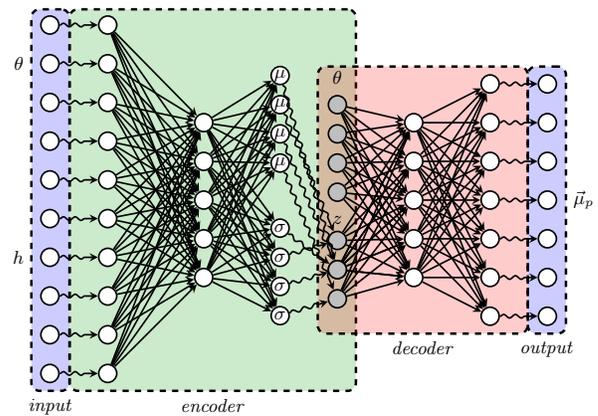
\begin{figure}[t]
\resizebox{0.45\textwidth}{!}{
\begin{tikzpicture}
	\node (1) [draw, dashed, minimum height=14em, minimum width=2em, xshift=23em, yshift=3em, fill=blue, fill opacity=0.2, very thick, rectangle, rounded corners] {};
	\node (la1) [below=0em of 1] {\emph{output}};
	\node (1) [draw, dashed, minimum height=20em, minimum width=15em, xshift=5.5em, yshift=3em, fill=olivegreen, fill opacity=0.2, very thick, rectangle, rounded corners] {};
	\node (la1) [below=0em of 1] {\emph{encoder}};
	\node (2) [draw, dashed, minimum height=14em, fill = red, fill opacity=0.2,minimum width=11em, xshift=16.5em, yshift=3em, very thick, rectangle, rounded corners] {};
	\node (la1) [below=0em of 2] {\emph{decoder}};
	\node (3) [draw, dashed, minimum height=20em, fill = blue, fill opacity=0.2,minimum width=2em, xshift=-3em, yshift=3em , very thick, rectangle, rounded corners] {};
		\node (la3) [below=0em of 3] {\emph{input}};
	
	\node[circle, thick, fill=white!50, draw] (x1) {};
	\node[circle, thick, draw, fill=white!50, below=1em of x1] (x2) {};
	\node[circle, thick, fill=white!50, draw, below=1em of x2] (x3) {};
	\node[circle, thick, fill=white!50, draw, below=1em of x3] (x4) {};
	\node[circle, thick, fill=white!50, draw, above=1em of x1] (x5) {};
	\node[circle, thick, fill=white!50, draw, above=1em of x5] (x6) {};
	\node[circle, thick, fill=white!50, draw, above=1em of x6] (x7) {};
	\node[circle, thick, fill=white!50, draw, above=1em of x7] (x8) {};
	\node[circle, thick, fill=white!50, draw, above=1em of x8] (x9) {};
	\node[circle, thick, fill=white!50, draw, above=1em of x9] (x10) {};

	\node[circle, thick, fill=white, left=2em of x1, draw] (i1) {};
	\node[circle, thick, draw, fill=white, below=1em of i1] (i2) {};
	\node[circle, thick, fill=white, draw, below=1em of i2] (i3) {};
	\node[circle, thick, fill=white, draw, below=1em of i3] (i4) {};
	\node[circle, thick, fill=white, draw, above=1em of i1] (i5) {};
	\node[circle, thick, fill=white, draw, above=1em of i5] (i6) {};
	\node[circle, thick, fill=white, draw, above=1em of i6] (i7) {};
	\node[circle, thick, fill=white, draw, above=1em of i7] (i8) {};
	\node[circle, thick, fill=white, draw, above=1em of i8] (i9) {};
	\node[circle, thick, fill=white, draw, above=1em of i9] (i10) {};
	
	\foreach \x in {1,...,10}
		\draw[-stealth, decoration={snake, pre length=0.01mm, segment length=2mm, amplitude=0.3mm, post length=1.5mm}, decorate, thick] (i\x) -- (x\x);
	
	\node[circle, thick, right=4em of x1,yshift=3em, fill=white, draw] (xh1) {};
	\node[circle, thick, draw, fill=white, below=1em of xh1] (xh2) {};
	\node[circle, thick, fill=white, draw, below=1em of xh2] (xh3) {};
	\node[circle, thick, fill=white, draw, above=1em of xh1] (xh4) {};
	\node[circle, thick, fill=white, draw, above=1em of xh4] (xh5) {};
	\node[circle, thick, fill=white, draw, right=8em of x1, yshift=8em] (hm1) {};
	\node[circle, thick, draw, fill=white, below=0.5em of hm1] (hm2) {};
	\node[circle, thick, draw, fill=white, below=0.5em of hm2] (hm3) {};
	\node[circle, thick, draw, fill=white, above=0.5em of hm1] (hm4) {};
	\node[circle, thick, fill=white, draw, right=8em of x1, yshift=0em] (hs1) {};
	\node[circle, thick, draw, fill=white, below=0.5em of hs1] (hs2) {};
	\node[circle, thick, draw, fill=white, below=0.5em of hs2] (hs3) {};
	\node[circle, thick, draw, fill=white, above=0.5em of hs1] (hs4) {};
	\node[] at (hm1) (mu1) {$\mu$};
	\node[] at (hm2) (mu2) {$\mu$};
	\node[] at (hm3) (mu3) {$\mu$};
	\node[] at (hm4) (mu4) {$\mu$};
	\node[] at (hs1) (s1) {$\sigma$};
	\node[] at (hs2) (s2) {$\sigma$};
	\node[] at (hs3) (s3) {$\sigma$};
	\node[] at (hs4) (s4) {$\sigma$};

	\node[circle, thick, fill=lightgray, draw, right=11em of x1, yshift=8em] (a1) {};
	\node[circle, thick, draw, fill=lightgray, below=0.5em of a1] (a2) {};
	\node[circle, thick, draw, fill=lightgray, below=0.5em of a2] (a3) {};
	\node[circle, thick, draw, fill=lightgray, below=0.5em of a3] (a4) {};
	\node[circle, thick, fill=lightgray, draw, below=1.5em of a4] (h1) {};
	\node[circle, thick, draw, fill=lightgray, below=0.5em of h1] (h2) {};
	\node[circle, thick, draw, fill=lightgray, below=0.5em of h2] (h3) {};
	\node[circle, thick, draw, fill=lightgray, above=0.5em of h3] (h4) {};
	\node[circle, thick, right=15em of x1, yshift=3em, fill=white, draw] (oh1) {};
	\node[circle, thick, draw, fill=white, below=1em of oh1] (oh2) {};
	\node[circle, thick, fill=white, draw, below=1em of oh2] (oh3) {};
	\node[circle, thick, fill=white, draw, above=1em of oh1] (oh4) {};
	\node[circle, thick, fill=white, draw, above=1em of oh4] (oh5) {};
	\node[circle, thick, draw, fill=white, right=19em of x1,yshift=3em] (o1) {};
	\node[circle, thick, draw, fill=white, below=1em of o1] (o2) {};
	\node[circle, thick, draw, fill=white, below=1em of o2] (o3) {};
	\node[circle, thick, draw, fill=white, below=1em of o3] (o4) {};
	\node[circle, thick, draw, fill=white, above=1em of o1] (o5) {};
	\node[circle, thick, draw, fill=white, above=1em of o5] (o6) {};
	\node[circle, thick, draw, fill=white, above=1em of o6] (o7) {};
	\node[circle, thick, draw, fill=white, right=22em of x1,yshift=3em] (oo1) {};
	\node[circle, thick, draw, fill=white, below=1em of oo1] (oo2) {};
	\node[circle, thick, draw, fill=white, below=1em of oo2] (oo3) {};
	\node[circle, thick, draw, fill=white, below=1em of oo3] (oo4) {};
	\node[circle, thick, draw, fill=white, above=1em of oo1] (oo5) {};
	\node[circle, thick, draw, fill=white, above=1em of oo5] (oo6) {};
	\node[circle, thick, draw, fill=white, above=1em of oo6] (oo7) {};

	\foreach \x in {1,...,10}
		\foreach \y in {1,...,5}
			\draw[-stealth, thick] (x\x) -- (xh\y);
	
	\foreach \x in {1,...,5}
		\foreach \y in {1,...,4}
			\draw[-stealth, thick] (xh\x) -- (hm\y);
	
	\foreach \x in {1,...,5}
		\foreach \y in {1,...,4}
			\draw[-stealth, thick] (xh\x) -- (hs\y);
	
	\foreach \x in {1,...,4}
		\draw[-stealth, decoration={snake, pre length=0.01mm, segment length=2mm, amplitude=0.3mm, post length=1.5mm}, decorate, thick] (hs\x) -- (h\x);
	\foreach \x in {1,...,4}
		\draw[-stealth, decoration={snake, pre length=0.01mm, segment length=2mm, amplitude=0.3mm, post length=1.5mm}, decorate, thick] (hm\x) -- (h\x);
	
	\foreach \x in {1,...,5}
		\foreach \y in {1,...,4}
			\draw[-stealth, thick] (h\y) -- (oh\x);
			
	\foreach \x in {1,...,5}
		\foreach \y in {1,...,4}
			\draw[-stealth, thick] (a\y) -- (oh\x);
	
	\foreach \x in {1,...,5}
		\foreach \y in {1,...,7}
			\draw[-stealth, thick] (oh\x) -- (o\y);
				
	\foreach \x in {1,...,7}
		\draw[-stealth, decoration={snake, pre length=0.01mm, segment length=2mm, amplitude=0.3mm, post length=1.5mm}, decorate, thick] (o\x) -- (oo\x);
				
	\node[left=0.5em of i1] (l1) {$h$};
	\node[left=0.5em of i9] (l1) {$\theta$};	
	\node[above=0em of h1] (l2) {$z$};
	\node[above=0.25em of a1] (l2) {$\theta$};
	\node[right=0.5em of oo1] (l3) {$\vec{\mu}_p$};

\end{tikzpicture}
}
\caption{Diagram of a CVAE. The white circles of the image represent
  the neurons of the neural network, which are nonlinear functions,
  while the connection between the neurons are represented by the
  black arrows. The CVAE in the diagram is broken up into four pieces:
  the input, the encoder, the decoder, and the output. The information
  flows through the CVAE from left to right and is compressed in the
  latent space. At the intersection of the encoder and decoder, we
  have the latent space, which consists of the latent variables $z$
  and the conditional variables $\theta$ where $\dim(z)<\dim(h)$. 
   The last layer of the encoder outputs $\vec{\mu}_q, \vec{\sigma}_q$ which 
   are used to  define a normal distribution $\mathcal{N}(\vec{\mu}_q,
   \vec{\sigma}_q)$. The latent layer then samples from this distribution, $z \sim q(z|h,\theta) =\mathcal{N}(\vec\mu_q(h,\theta), \vec\sigma_q(h,\theta))$. The  last layer of the encoder consists the $\vec\mu_q(h,\theta)$ and $\vec\sigma_q(h,\theta)$
  variables which are, respectively, used as the mean and standard
  deviation of the normal distribution sampled from the latent
  variables.  These variables facilitate interpolation by forcing the
  latent variables to be normally distributed.  
\label{CVAE}
}
\end{figure}

The loss \eqref{eq:L} consists of two terms. By minimizing the first
term (the reconstruction loss) the CVAE attempts to reconstruct $h$ as
well as possible after first being encoded by $q(z|h,\theta)$ into a
latent representation, and then decoded by $p(h|z,\theta)$ into a new
waveform. In this sense, the CVAE is similar to a vanilla autoencoder
(see, e.g.,~\cite{Goodfellow-et-al-2016}). The second term (the KL
loss) pushes $q(z|h,\theta)$ to match the prior $p(z)$. In this way,
it regularizes the model, discouraging $q(z|h,\theta)$ from memorizing
the training data. See Fig.~\ref{CVAE} for an illustration of the
overall structure.

\begin{figure}[ht]
\begin{tikzpicture}

	\node (1) [draw, dashed, minimum height=14em, minimum width=2em, xshift=23em, fill=blue, fill opacity=0.2, very thick, rectangle, rounded corners] {};
	\node (la1) [below=0em of 1] {\emph{output}};
	\node (2) [draw, dashed, minimum height=14em, fill = red, fill opacity=0.2,minimum width=11em, xshift=16.5em, very thick, rectangle, rounded corners] {};
	\node (la1) [below=0em of 2] {\emph{decoder}};

	\node[circle, thick, fill=lightgray, draw, right=11em of x1, yshift=4.5em] (a1) {};
	\node[circle, thick, draw, fill=lightgray, below=0.5em of a1] (a2) {};
	\node[circle, thick, draw, fill=lightgray, below=0.5em of a2] (a3) {};
	\node[circle, thick, draw, fill=lightgray, below=0.5em of a3] (a4) {};
	\node[circle, thick, fill=lightgray, draw, below=1em of a4] (h1) {};
	\node[circle, thick, draw, fill=lightgray, below=0.5em of h1] (h2) {};
	\node[circle, thick, draw, fill=lightgray, below=0.5em of h2] (h3) {};
	\node[circle, thick, draw, fill=lightgray, above=0.5em of h3] (h4) {};
	\node[circle, thick, right=15em of x1, fill=white, draw] (oh1) {};
	\node[circle, thick, draw, fill=white, below=1em of oh1] (oh2) {};
	\node[circle, thick, fill=white, draw, below=1em of oh2] (oh3) {};
	\node[circle, thick, fill=white, draw, above=1em of oh1] (oh4) {};
	\node[circle, thick, fill=white, draw, above=1em of oh4] (oh5) {};
	\node[circle, thick, draw, fill=white, right=19em of x1] (o1) {};
	\node[circle, thick, draw, fill=white, below=1em of o1] (o2) {};
	\node[circle, thick, draw, fill=white, below=1em of o2] (o3) {};
	\node[circle, thick, draw, fill=white, below=1em of o3] (o4) {};
	\node[circle, thick, draw, fill=white, above=1em of o1] (o5) {};
	\node[circle, thick, draw, fill=white, above=1em of o5] (o6) {};
	\node[circle, thick, draw, fill=white, above=1em of o6] (o7) {};
	\node[circle, thick, draw, fill=white, right=22em of x1] (oo1) {};
	\node[circle, thick, draw, fill=white, below=1em of oo1] (oo2) {};
	\node[circle, thick, draw, fill=white, below=1em of oo2] (oo3) {};
	\node[circle, thick, draw, fill=white, below=1em of oo3] (oo4) {};
	\node[circle, thick, draw, fill=white, above=1em of oo1] (oo5) {};
	\node[circle, thick, draw, fill=white, above=1em of oo5] (oo6) {};
	\node[circle, thick, draw, fill=white, above=1em of oo6] (oo7) {};

	\foreach \x in {1,...,5}
		\foreach \y in {1,...,4}
			\draw[-stealth, thick] (h\y) -- (oh\x);
			
	\foreach \x in {1,...,5}
		\foreach \y in {1,...,4}
			\draw[-stealth, thick] (a\y) -- (oh\x);
	
	\foreach \x in {1,...,5}
		\foreach \y in {1,...,7}
			\draw[-stealth, thick] (oh\x) -- (o\y);
				
	\foreach \x in {1,...,7}
		\draw[-stealth, decoration={snake, pre length=0.01mm, segment length=2mm, amplitude=0.3mm, post length=1.5mm}, decorate, thick] (o\x) -- (oo\x);
				
	\node[above=0em of h1] (l2) {$z$};
	\node[above=0.25em of a1] (l2) {$\theta$};
	\node[right=0.5em of oo1] (l3) {$\vec{\mu}_p$};

\end{tikzpicture}
\caption{After the CVAE is trained, one can disregard the 
input and the encoder and only consider the layers after the 
latent space. Doing so results in a model $\vec{\mu}_p = 
{\rm Dec}(\theta,z)$ where $\rm Dec$ is the decoder network.
\label{model}
}
\end{figure}
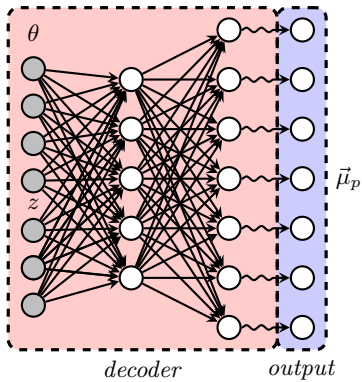

Following standard practice, we take all distributions over the latent space to be multivariate normal of dimension $l=\dim(z)$, with diagonal covariance,
\begin{align}
  q(z|h,\theta) &= \mathcal{N}(\vec\mu_q(h,\theta), \vec\sigma_q(h,\theta)), \label{eq:encoder}\\
  p(h|z,\theta) &= \mathcal{N}(\vec\mu_p(z,\theta), \vec 1), \label{eq:decoder} \\
  p(z) &= \mathcal{N}(0,1)^l.
\end{align}
The functions $(\vec\mu_q(h,\theta), \vec\sigma_q(h,\theta))$ defining
the mean and variance of the encoder are given as outputs of an
encoder neural network, which takes as input $(h,\theta)$.  
Likewise, the decoder mean
$\vec\mu_p(z,\theta)$ is given as the output of a decoder network, which
takes as input $(z,\theta)$. The covariance of the decoder is fixed to
the identity, and, following standard practice, the prior over $z$ is
taken to be a standard multivariate normal distribution. 

When written in the form \eqref{eq:L}, the loss $L$ may be evaluated,
and the neural networks trained. Since $q(z|h,\theta)$ and $p(z)$ are
both multivariate normal, the KL divergence can be evaluated
analytically for any $(h,\theta)$. The reconstruction loss is evaluated
using a single-sample Monte Carlo approximation as follows:
\begin{enumerate}
\item Use the encoder network to evaluate
  $(\vec\mu_q(h,\theta), \vec\sigma_q(h,\theta))$.
\item Sample $z\sim q(z|h,\theta)$ by first sampling
  $\vec \epsilon \sim \mathcal{N}(0,1)^l$ and then setting
  $z=\vec \mu_q(h,\theta) + \vec\epsilon \odot
  \vec\sigma_q(h,\theta)$. This is known as the
  \emph{reparameterization trick}~\cite{kingma2013autoencoding}.
\item Use the decoder network to evaluate $\vec\mu_p(z,\theta)$.
\item Evaluate the loss $-\log p(h|z,\theta)$.
\end{enumerate}
The loss $L$ and its derivatives with respect to neural network
parameters may then be evaluated on minibatches and minimized using a
standard gradient-based stochastic optimizer (we use
Adam~\cite{kingma2017adam}). The reparameterization trick was used to
separate out the stochastic aspect of sampling from $q(z|h,\theta)$
and enable backpropagation for calculating the derivatives.

We now comment on the relation of this procedure to standard fitting
techniques for model building. If $p(h|z,\theta)$ were independent of
$z$, then the reconstruction loss would no longer depend on
$q(z|h,\theta)$. For the multivariate normal $p(h|\theta)$ given in
Eq.~\eqref{eq:decoder} (now omitting $z$) this reduces (up to
normalization) to the mean squared difference
$\frac{1}{N}\sum_{i=1}^N\|\vec h^{(i)} - \vec
\mu_p(\theta^{(i)})\|^2$, i.e., $\vec\mu_p(\theta)$ becomes a model
for the waveform given by a standard least-squares fit to training
data. (One could additionally modify the covariance of $p(h|\theta)$,
to specialize the fit, e.g., by including information about detector
noise properties, or to allow it to be fit as well during training.)

By allowing for dependence on the latent variables $z$, the
marginalized distribution $p(h|\theta)$ can have a more complicated
(non-Gaussian) structure. This is needed to build a model from
simulations that depend on parameters not included in $\theta$, as
these could give rise to different $h$ for the same $\theta$. In the
following sections, using a toy model for BNS
postmerger waveforms, we provide evidence that the latent variables
can learn about the hidden variables of the training data, in this
case the EOS. Given a gravitational-wave detection, by using
$p(h|z,\theta)$ to perform Bayesian parameter estimation jointly over
$\theta$ and $z$, one could, therefore, learn useful information about
which of these parameters are preferred; see Fig.~\ref{model}.

The fact that one optimizes $L$ rather than $L_{\text{MLL}}$ when
training a CVAE gives rise to a possible pitfall. Notice that if the
last term in Eq.~\eqref{eq:Lcompare} vanishes (i.e., if
$q(z|h,\theta)$ and $p(z|h,\theta)$ are identical) then the two losses
coincide. Typically, though, the form~\eqref{eq:encoder} of the
encoder $q(z|h,\theta)$ is too restrictive to properly represent the
posterior $p(z|h,\theta)$, so $L > L_{\text{MLL}}$. Another way
equality can be achieved, however, is by ignoring the latent space
entirely, i.e., by setting $p(h|z,\theta) = p(h|\theta)$ and
$q(z|h,\theta) = p(z|h,\theta) = p(z)$. This is known as posterior
collapse. In other words, optimizing $L$ rather than $L_{\text{MLL}}$
means that the use of the latent space incurs a cost. If this cost
exceeds the gain in reconstruction performance achieved by using the
latent variables, then the CVAE can fail to
autoencode~\cite{chen2017variational}. To mitigate this problem, one
approach is to multiply the KL loss term by a factor
$\beta < 1$~\cite{higgins2016beta}, and possibly anneal this term to
one during training~\cite{bowman2016generating}. In this work, we
determine experimentally a fixed value for $\beta$ that gives good
performance. Other approaches include generalizing the form of the
encoder distribution so that it more easily approximates the
posterior~\cite{kingma2017improving}. In Appendix~\ref{sec:toy_model},
we demonstrate these methods on a toy model where the data is a simple
sum of sine waves.

\section{Postmerger waveform model}\label{sec:imp}

In this section, we apply the CVAE framework to construct a model for
postmerger waveforms based on numerical relativity simulations. We
build a simplified model $p(h|M,z)$, taking the parameter space to be
the total mass of the system, i.e.,~$\theta = M$, and representing
waveforms in terms of several phenomenological fit parameters. We find
that the small number of available numerical relativity simulations
leads to overfitting, so in the following section we assess the CVAE
approach further using synthetic training data.

\subsection{Training data}

We obtain numerical training waveforms from the CoRe database
\cite{dietrich2018core}. This comprises 367 BNS waveforms, with
varying component masses, spins, and equations of state, as well as
different numerical resolutions and starting frequencies. (There are
163 unique choices of masses, spins, and equations of state.) From
these inspiral-merger-ringdown waveforms, we extract the postmerger
signal by identifying the moment in time of peak gravitational-wave
amplitude, and truncating the preceding signal. 
Note that although the neutron stars come into contact prior to this peak, this
definition of the postmerger signal is sufficient for our purposes.  We drop
all waveforms that exhibit prompt collapse to a black hole, requiring that the
post-peak waveform lasts at least 5~ms.
 
The CoRe simulations have been performed at different resolutions and
for varying time durations. To prepare the waveform data for the
neural network, we standardize the time resolution and total time by
resampling using a quadratic spline and padding with zeros at the
end. We represent waveforms using 1000 time samples, with a total
duration of 45~ms, which is the duration of the longest postmerger
signal. Finally, for simplicity, we only take the real part of the
$\ell=m=2$ component of the complex strain $h = h_+ - i h_\times$,
after multiplying by a phase such that the real part vanishes at the
merger time, $t=0$, i.e.,~
$h(t) \equiv \mathrm{Re}[e^{i\theta}h_{22}(t)]$, with $h(t=0)=0$.

When fitting the model directly to these strain data sets, we find
that the zero padding leads to poor waveform reconstruction with
overdamping at late times (see App.~\ref{FittingTime}). Instead, we
find it more effective to use a compressed representation, where
waveforms are expressed in terms of parameters of a phenomenological
fit~\cite{Bose_2018},
\begin{align}
\label{rezz}
 h(t) ={}& \alpha_1 e^{-t/\tau_1}\left\{\sin\left(2\pi f_1 t\right) +\sin\left[2\pi(f_1-f_{1\epsilon})t\right]\right.\nonumber \\
           & \qquad \qquad  \left. +\sin\left[2\pi(f_1+f_{1\epsilon})t\right]\right\} \nonumber \\
           & + \alpha_2 e^{-t/\tau_2}\sin\left(2\pi f_2 t + 2\pi \gamma_2 t^2 \right. \nonumber \\
           & \qquad \qquad \qquad \quad \left. +2\pi \xi_2 t^3 + \pi \beta_2\right).
\end{align}
Here, $f_{1\epsilon}=50$ Hz, and the parameters
$\{\alpha_1,\alpha_2,\tau_1,\tau_2,f_1,f_2,\beta_2,\gamma_2,\xi_2\}$
are determined by least-squares fit. This model is motivated by the
fact that BNS postmerger signals tend to consist of two damped
sinusoids, with corrections added for improved
accuracy~\cite{Bose_2018}.

The parameterized form \eqref{rezz} is capable of representing the
numerical-relativity postmerger signals with $\mathcal M > 0.8$, where
\begin{equation}
\label{match}
\mathcal{M}(h,\tilde{h}) = \frac{\langle h|\tilde{h}\rangle}{\sqrt[]{\langle h | h \rangle \langle \tilde{h} | \tilde{h} \rangle}}, \quad
\langle h | \tilde{h} \rangle = \int_{0}^{t_f} h(t)\tilde{h}^*(t)dt,
\end{equation}
is the match between two waveforms, assuming a flat noise spectrum. We
additionally restrict to those waveforms with $\mathcal M >0.9$. This
reduces our data set to 123 waveforms\footnote{The vast majority of 
these cases are low eccentricity quasi-circular inspirals, though
we do include three cases with eccentricity in the range $0.04$--$0.15$
that we have explicitly checked do not give rise to outliers in terms
of the best fit coefficients of Eq.~\eqref{rezz}.}. Finally, we use principle
component analysis whitening to decorrelate and standardize the
waveform parameters (see App.~\ref{appendix data process}).

\begin{figure*}[tb]
	\includegraphics[width=0.3\textwidth]{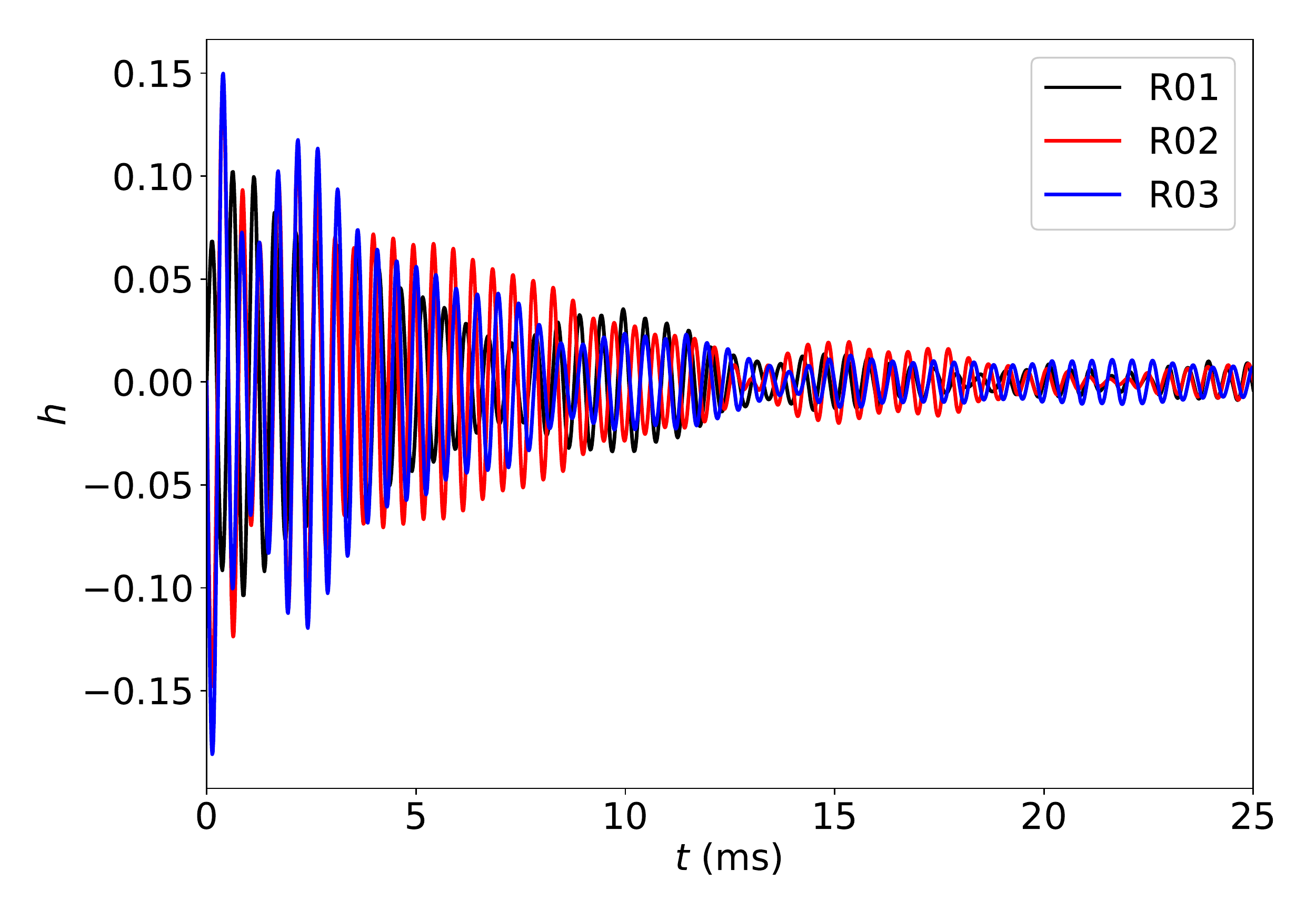}
    \includegraphics[width=0.3\textwidth]{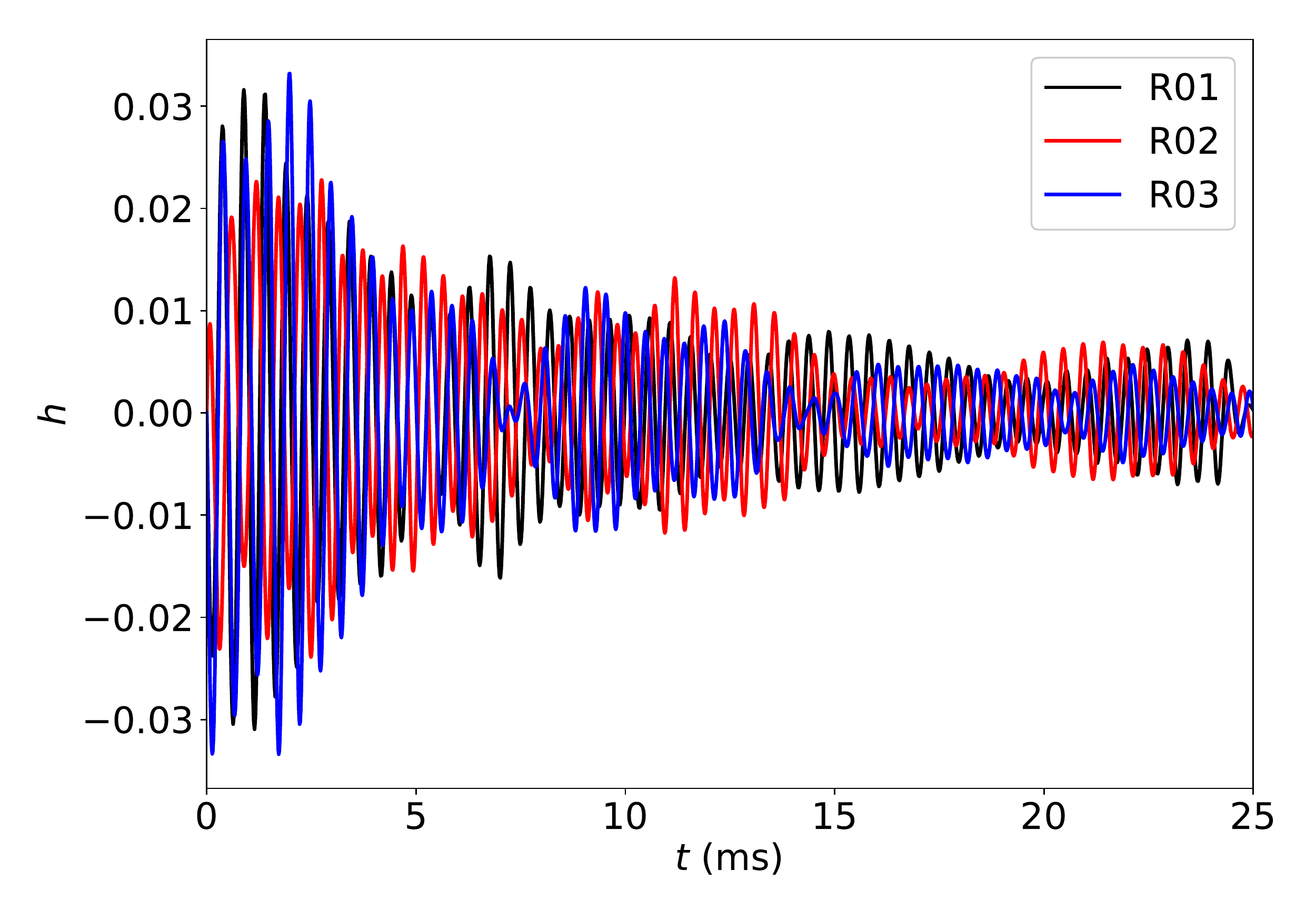}
	\includegraphics[width=0.3\textwidth]{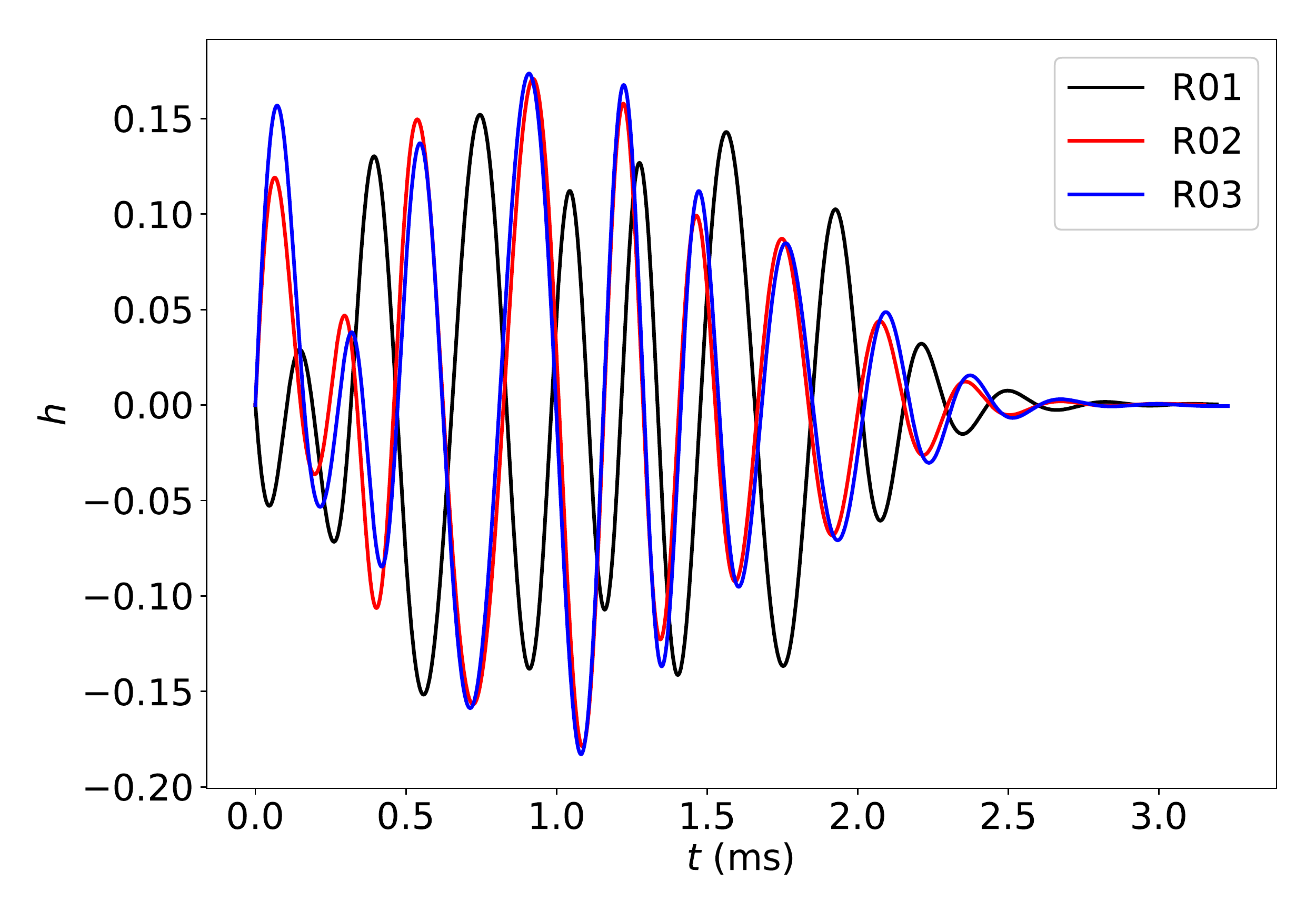}
	\caption{
        Post merger signal $h$ from simulations from the CoRe database
        performed at three different resolutions.  From left to right, the
        panels respectively correspond to simulations with mass-ratios $q=1.0$, $0.5$, and $1.0$; total masses of
        $2.75 M_\odot$, $2.88 M_\odot$, and $2.70 M_\odot$; and use the MS1b (left and middle) and SLy (right panel) EOSs. 
        (They are respectively labelled BAM0070,
        BAM0094, and BAM0100 in the CoRe database, see Ref.~\cite{dietrich2018core} for more details.) The black, red, and blue curves in each panel
        correspond to simulations with the same physical parameters, but with
        increasing numerical resolution. 
        From left to right, computing the match between low and high resolution (medium and high resolution) for each case, 
        we obtain, respectively, 0.11, 0.52, and 0.20 (0.12, 0.56, and 0.95).
	\label{fig:resolution_comp}
    }
\end{figure*}

We emphasize that the numerical uncertainties in the waveforms, at least at the
level of the phase, are much larger than the fitting error. In
Fig.~\ref{fig:resolution_comp}, we show three example resolution studies of the
post-merger waveforms. The curves in each panel in
Fig.~\ref{fig:resolution_comp} represent the same physical system, but quickly
become out of phase with each other leading to a poor match between them.
Computing the match between waveforms at different resolutions by picking the
post-merger part of the signal as described above and computing the match
between the highest and lowest resolution simulations, we find matches in the
range 0.11--0.52 for the cases shown.  This is before even taking into account
uncertainties regarding the EOS and other microphysics that is not included in
the simulations.

To summarize, our training data consist of 123 $(M, h)$ pairs, where
$h$ is a 9-component vector describing a postmerger signal. This
discards labels for the mass ratio, spins, EOS, and numerical
resolutions, but keeps that of the total mass. This omitted
information, however, gives rise to differences between the waveforms,
so it (along with any other differences between simulations) should be
understood as latent variables. In the following subsection, we
construct a CVAE to model these waveforms and characterize the latent
information.

\subsection{Experiments}
\label{sec:Exp}
We model $p(h|M)$ with a CVAE with latent dimension $l=4$. The
encoder and decoder are fully-connected networks with rectified
linear unit nonlinearities. Exploring various hyperparameter choices,
we find best performance with a 2-hidden-layer
encoder with $(100, 50)$ units, a decoder with inverted
structure, and $\beta=0.003$. 

\begin{figure}[]
\includegraphics[width=0.45\textwidth]{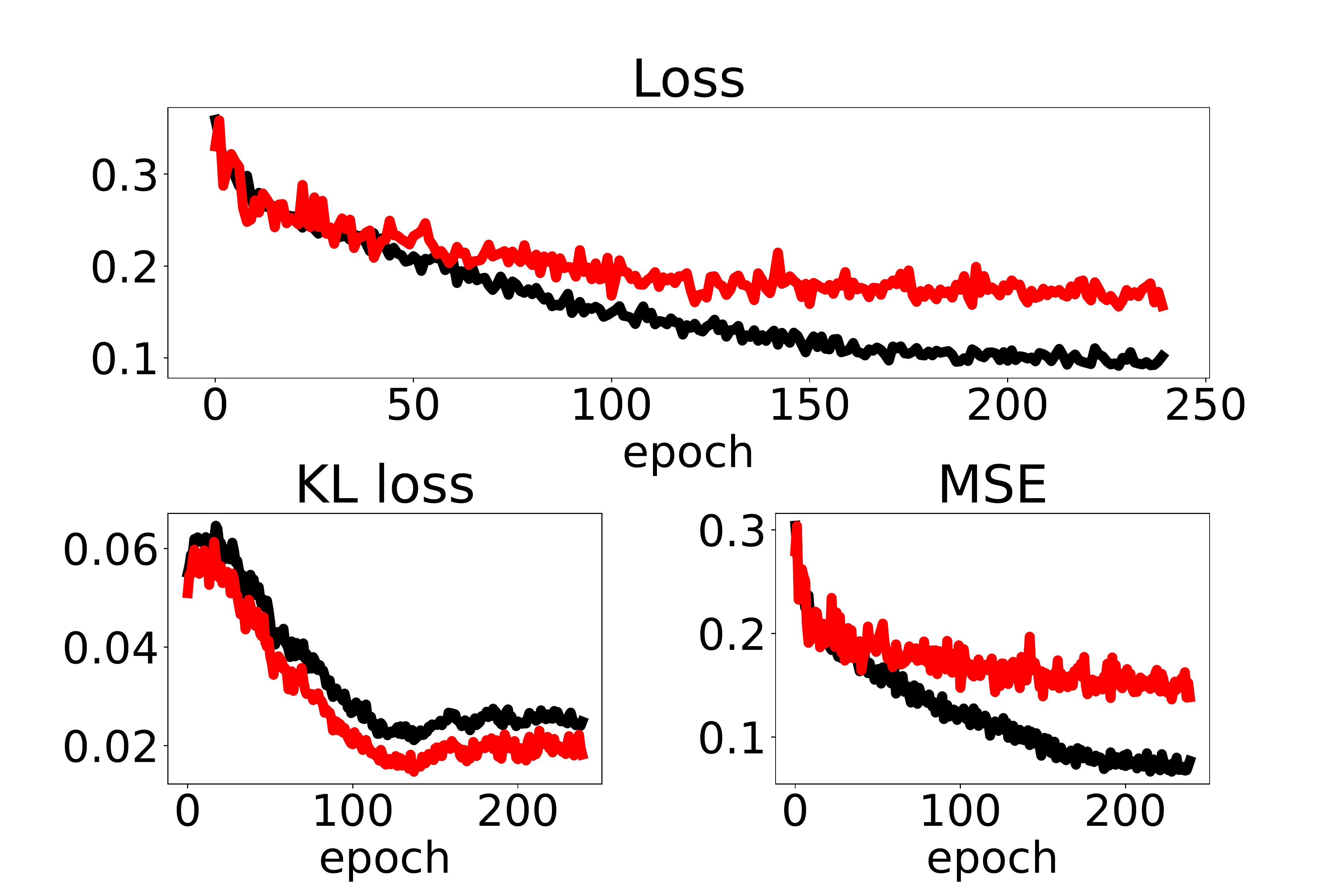}
\caption{Training history for a CVAE trained with numerical waveforms
  in the representation \eqref{rezz}. The upper plot shows the
  total loss \eqref{eq:L}, while the bottom two plots show the KL
  (left) and reconstruction loss (right) contributions to the
  total. The black and red curves correspond to the training and
  validation sets, respectively. 
  The gap between training and
  validation sets is an indication of overfitting.}
\label{train}
\end{figure}

We split our dataset into training (60\%) and validation (40\%) sets,
and train for 4000 epochs. 
The training history is plotted in
Fig.~\ref{train}. Since the validation loss is much higher than the
training loss, we conclude that the CVAE is overfitting, and that the
training set is too small for the network to successfully generalize
to previously unseen waveforms. We have also tried methods for mitigating overfitting such as adding dropout layers. There was no significant improvements using that method.

\section{Extended training set}
\label{sec:extended}

\begin{figure*}[]
  \includegraphics[width=\textwidth,trim=50 80 100 80]{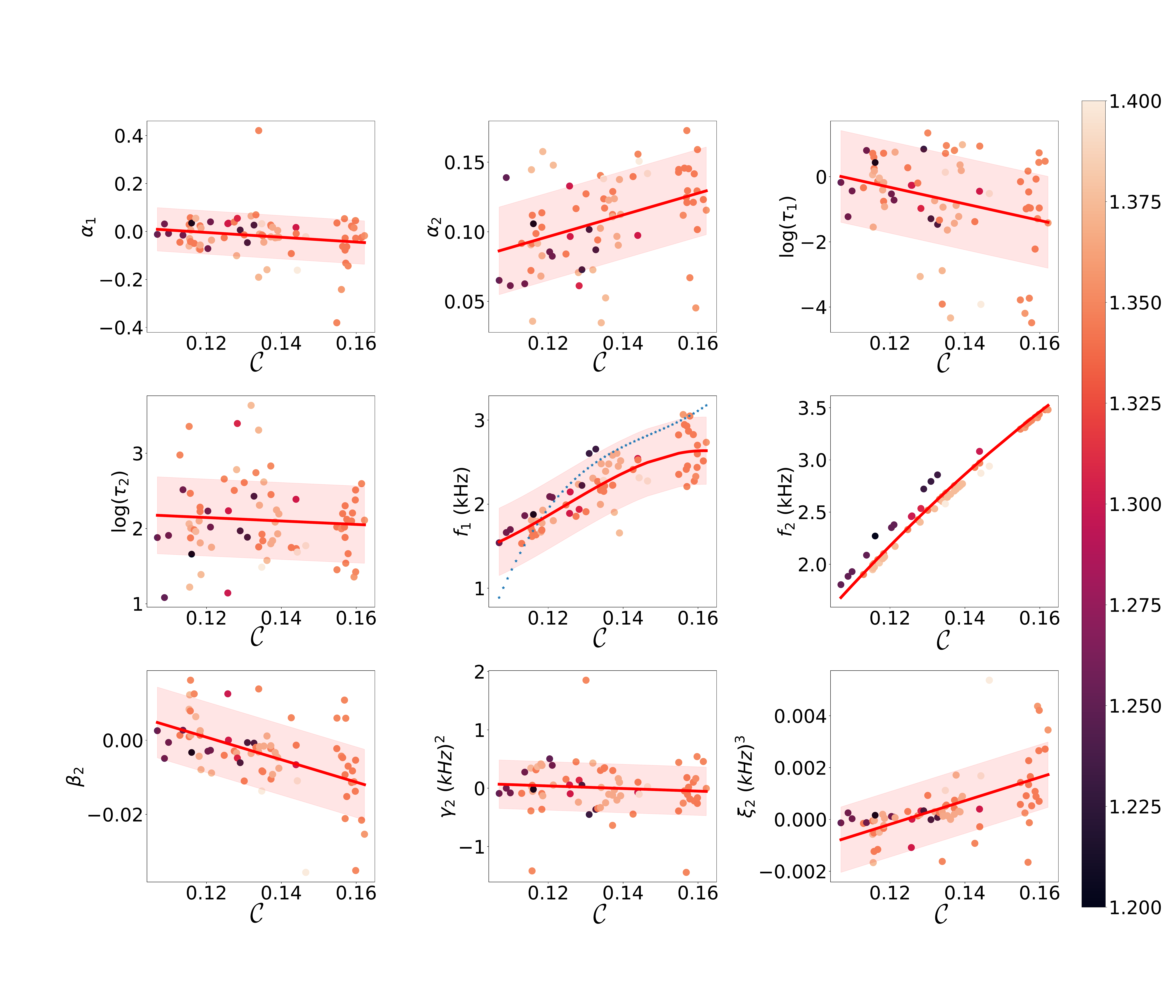}
  \caption{The parameters in Eq.~(\ref{rezz}), obtained from fitting
    the numerical waveforms, versus compactness $ \mathcal{C}$, color
    coded by the average mass $(\bar{M})$ of the two stars in units of
    solar mass.  Since we do not have direct access to $\mathcal{C}$
    from the simulations, here we define it be given by
    $f_2 = (a_9 +a_{10} \mathcal{C} + a_{11} \mathcal{C}^2)(1.6\ 
    M_\odot/\bar{M})$ with $a_9=-3.12$ kHz, $a_{10}=51.90$ kHz, and
    $a_{11}=-89.07$ kHz. We then fit for the parameters of
    $f_1 = (a_5 + a_6 \mathcal{C} + a_7 \mathcal{C}^2+ a_8
    \mathcal{C}^3)(1.6\ M_\odot/\bar{M})$ with our data. The blue curve
    in the $f_1$ plot represents the fit from
    Ref.~\cite{Bose_2018}. Though this curve roughly lies within the
    range of the data points, we find different values for the best
    fit parameters for the mean of $f_1$. The remaining parameters, on
    the other hand, appear to be roughly linear in $\mathcal{C}$, with
    some noise. We overlay the fiducial model at a fixed
    $\bar{M}=1.35\ M_{\odot}$ on the numerical data: the solid red
    line is the mean of the data, i.e.
    $\vec{A}\cdot \mathcal{C}+ \vec{B} $, while the shaded regions are
    the one standard deviation ranges.  From this, we see that the
    fiducial model captures most of the numerical data.  }
	\label{params}
\end{figure*}

The number of available numerical simulations is at present too small
to properly train a CVAE to model postmerger waveforms. Nevertheless,
we can assess the viability of the approach for the future by training
on synthetic data. In this section, we first construct a ``fiducial''
waveform model $p_{\text{fid}}(h|\bar M, \mathcal C)$ based on the
CoRe waveforms. The fiducial model depends on the average mass
$\bar{M} = (M_1+M_2)/2$ and the compactness $\mathcal{C} = GM/(Rc^2)$
of the remnant neutron star. We then generate a much larger training
set by sampling from this model, and we show that with this we can train the CVAE to have similar statistical properties as the fiducial model. This exercise also provides an estimate for the number of numerical waveforms needed to train a CVAE.

\subsection{Fiducial model}\label{sec:fm}

We now construct the fiducial model
$p_{\text{fid}}(h|\bar M, \mathcal C)$ for postmerger waveforms $h$
given $\bar{M}$ and $\mathcal{C}$. We do not have direct access to
$\mathcal{C}$ from the numerical-relativity simulations, so instead we
extract it based on the dominant postmerger
frequency~\cite{Lehner:2016lxy,Bose_2018,Rezzolla_2016}. Indeed,
Ref.~\cite{Bose_2018} found that, for a remnant with
$\bar{M} = 1.6~\mathrm{M}_{\odot}$, the dominant postmerger
frequencies approximately satisfy the relations,
\begin{align}
  \label{eq:f1}
  f_1 &= \left(a_5 +a_6 \mathcal{C} + a_7\mathcal{C}^2 + a_8\mathcal{C}^3 \right)(1.6\ M_\odot/\bar{M}),\\
  \label{eq:f2}
  f_2 &= \left( a_{9} +a_{10} \mathcal{C} + a_{11}\mathcal{C}^2 \right)(1.6\ M_\odot/\bar{M}) ,
\end{align}
with the numerical parameters $a_5 = -35.17$ kHz,
$a_6 = 727.99$ kHz, $a_7 = -4858.54$ kHz,
$a_8 = 10989.88$ kHz, $a_9=-3.12$ kHz,
$a_{10}=51.90$ kHz, and $a_{11}=-89.07$ kHz. 
On dimensionful grounds, we have added an extra factor $1/\bar{M}$ to  extend
these relations to other BNS masses.  We therefore take Eq.~\eqref{eq:f2} as
given and use it to \emph{define} $\mathcal{C}$ in terms of $f_2$ and $\bar M$.  
After solving for $\mathcal{C}$, we
plot all the parameters from fitting Eq.~(\ref{rezz}) to the numerical
waveforms versus $\mathcal{C}$ in Fig.~\ref{params}. The parameters of
$f_1$ are then fitted to our data. We find $a_5 = 11.26$
kHz, $a_6 = -284.83$ kHz, $a_7 = 2515.07$ kHz, and
$a_8 = -6800.26$ kHz. These values are different from the ones
found in Ref.~\cite{Bose_2018}, but as can be seen in Fig.~\ref{params}, the
curve defined by \cite{Bose_2018} is within one standard deviation of
the parameters we find.

We fit a generalized linear model for the remaining waveform
parameters in terms of the compactness, i.e., 
\begin{equation}
\label{fidModel}
\begin{pmatrix}
\alpha_1\\
\alpha_2\\
\log(\tau_1)\\
\log(\tau_2) \\
\frac{\bar{M}}{1.6M_\odot}f_1 \\
\frac{\bar{M}}{1.6M_\odot}f_2 \\
\beta_2 \\
\frac{\bar{M}^2}{1.6M_\odot}\gamma_2\\
\frac{\bar{M}^3}{1.6M_\odot} \xi_2\\
\end{pmatrix}
\sim \mathcal{N}(\vec B + \vec A \mathcal{C},  \Sigma) ,
\end{equation}
where
\begin{align}
\vec{A}&=\begin{pmatrix}
a_1\\
a_2\\
a_3\\
a_4\\
\left(a_5/\mathcal{C}+a_6+a_7\mathcal{C}+a_{8}\mathcal{C}^2\right)\\
\left(a_{9}/\mathcal{C} + a_{10} + a_{11} \mathcal{C}\right)\\
a_{12}\\
a_{13}\\
a_{14}
\end{pmatrix},
\end{align}
\begin{align}
\vec{B}&=\begin{pmatrix}
b_1\\
b_2\\
b_3\\
b_4\\
b_5\\
b_6\\
b_7\\
b_8\\
b_9
\end{pmatrix},
\end{align}
 and where $\Sigma$ is a generic
covariance matrix, and $a_i$ and $b_i$ are constants which are
obtained by fitting the numerical waveforms except for $\{a_5,a_6,a_7,a_8,a_9,a_{10},a_{11}\}$, where we use the values discussed previously. 
We
determine $\Sigma$ by first subtracting the mean of
$\{\alpha_1,\alpha_2,\log(\tau_1),\log(\tau_2), \beta_2
,\gamma_2,\xi_2\}$, and subtracting Eq.~(\ref{eq:f1}) and
Eq.~(\ref{eq:f2}) for the frequencies. This method of computing
$\Sigma$ is conservative in the sense that it overestimates the noise
in the model. The covariance matrix can also be estimated by first
removing $\vec{A}\cdot \mathcal{C} + \vec{B}$ from the data, or by
simultaneously fitting for the linear dependence and covariance.
However, we have checked that doing the former only reduces the
magnitude of the covariance $\sim2\%$, and does not significantly
affect the results of the estimate.

Note that we took logarithms of $\tau_1$ and $\tau_2$ so as to
normalize their distributions. The values of the parameters are
recorded in appendix \ref{paramvalues}. We add a factor of $1/\bar{M}$
in the frequencies to roughly model mass dependence of the frequency
on the mass, inspired by Ref.~\cite{Lehner:2016lxy}. The estimated
variance in $\gamma_2$ and $\xi_2$ is large, which can lead to issues with
the waveforms at late times. On dimensional grounds, we also add a
factor of $1/\bar{M}^2$ to $\gamma_2$ and $1/\bar{M}^3$ to
$\xi_2$. 

We underline the fact that due to the stochastic nature of this model, even fixing $\bar{M}$
and $\mathcal{C}$, we will obtain different waveforms with each realization.
Generating a large set of realizations of the fiducial model with these
quantities fixed, we find that average match between the waveforms is $0.38$.
These is similar to the values obtained when varying numerical resolutions in
the simulations shown in Fig.~\ref{fig:resolution_comp}, and thus representative of
the large theoretical uncertainties. Furthermore, when training a CVAE on this
fiducial model, we expect it to incorporate similar uncertainties: if the
CVAE is well trained, then the CVAE model could recreate any of the waveforms
from the fiducial model giving an average match close to $0.38$.

\subsection{Experiments}
\label{exp}

\begin{figure*}[]
  \includegraphics[width=\textwidth]{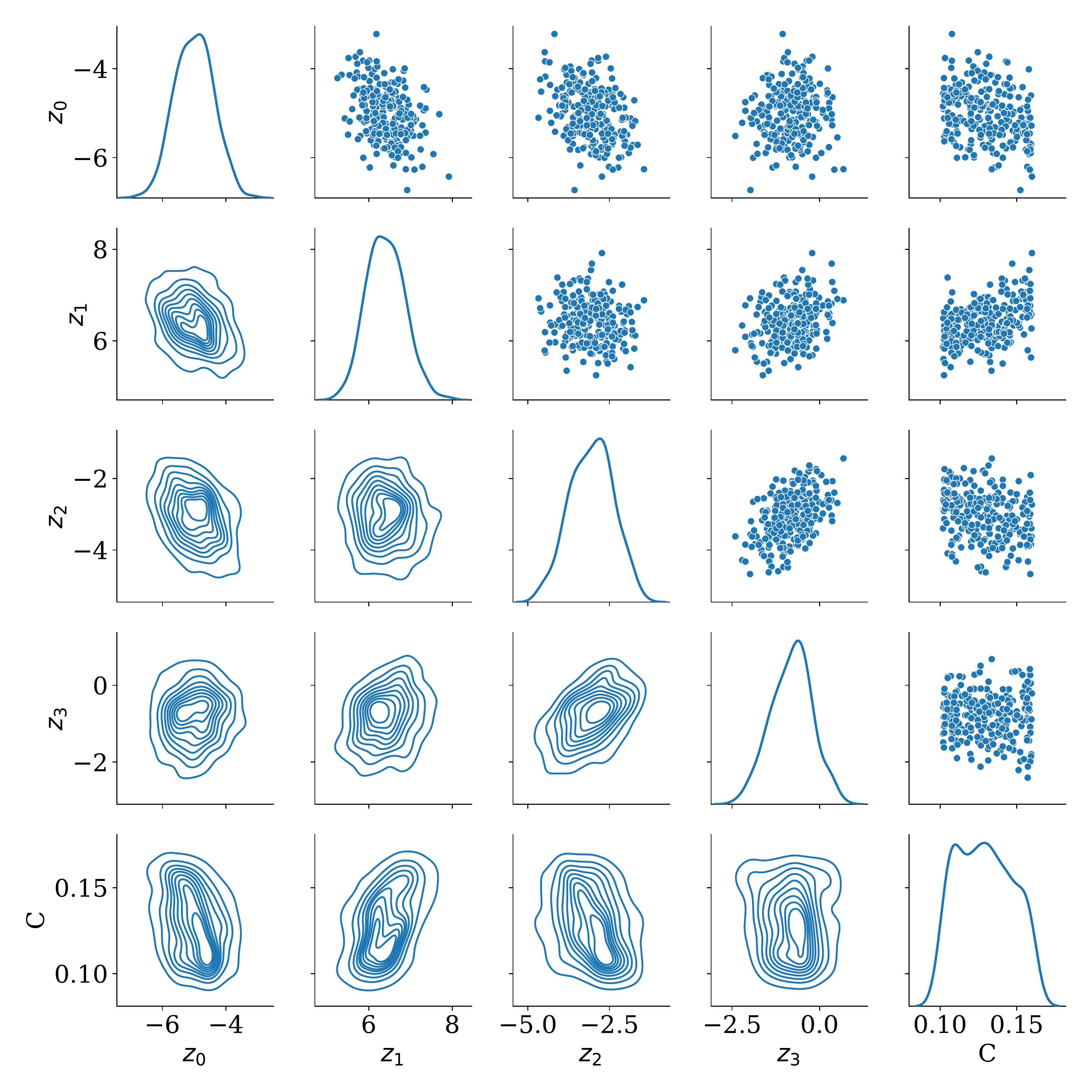}
  \caption{Pairplot of latent variables generated with data generated
    with the fiducial model with fixed masses and random
    compactness $\mathcal C \sim p(\mathcal C)$. That is, we sample $h\sim p_{\text{fid}}(h|\bar M =1.35M_\odot ,\mathcal C)$, then
    $z \sim p(z|\bar{M}=1.35
    M_\odot,h)$.  
    We can see that the
    CVAE is avoiding posterior collapse, which would be indicated by
    that data being independent of some latent variable. By looking at the compactness
    column one would hope to see a relationship between the
    compactness and the latent variables. Unfortunately the
    relationship is not obvious here.  }
  \label{z_1}
\end{figure*}

\begin{figure*}[]
  \includegraphics[width=\textwidth]{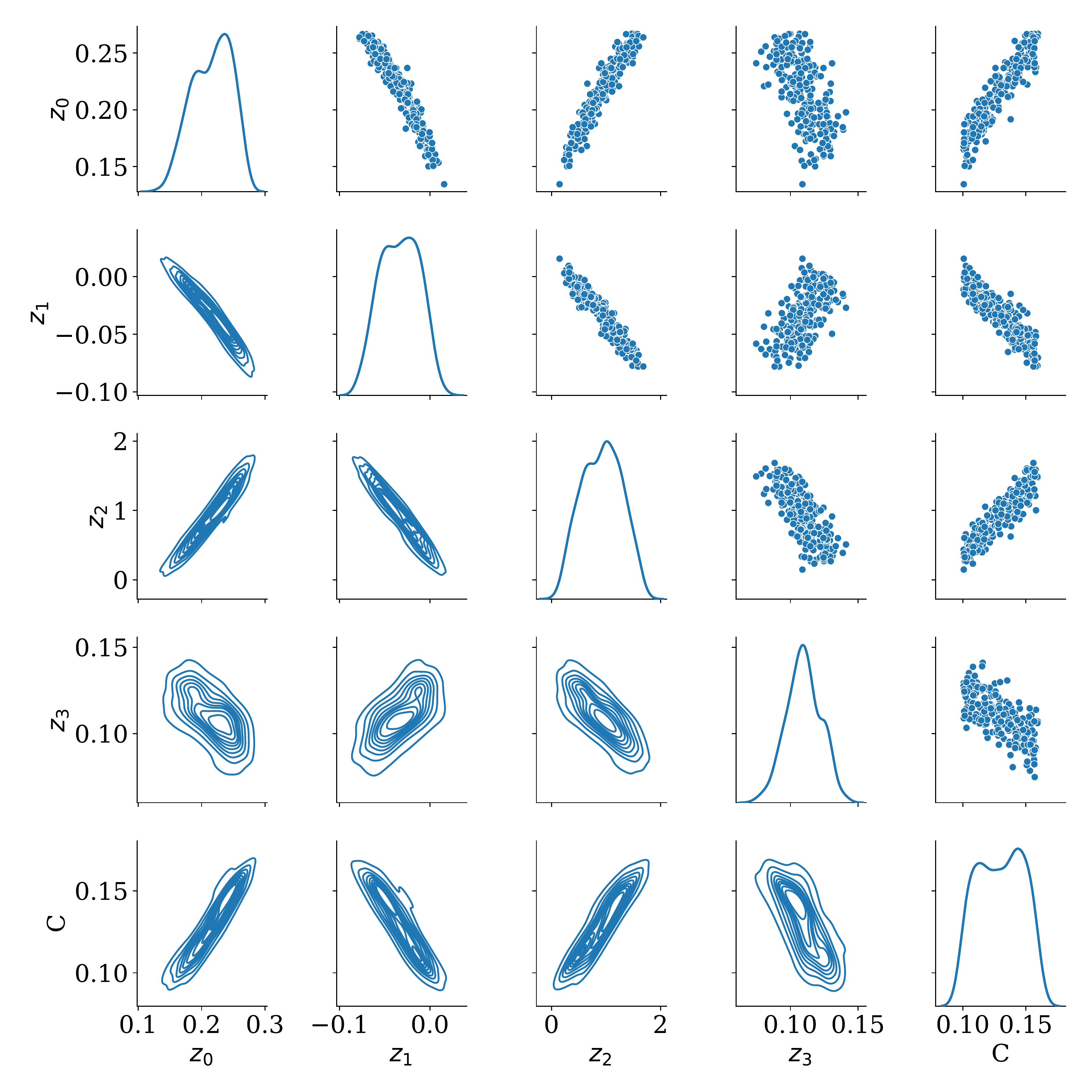}
  \caption{
    Similar to Fig.~\ref{z_1}, but for the latent space of the CVAE trained on a
    fiducial model where the variance is reduced by half, i.e., $\Sigma \rightarrow
    0.5\Sigma$.  % We show a pairplot of latent variables generated with fixed masses
    % and random compactness. That is, we sample $z \sim
    % p(z|\bar{M}=1.35M_\odot,h(\bar{M}=1.35M_\odot,\mathcal{C}))p(\mathcal{C})$.
    In
    this case, we can see that there is a roughly linear relationship between the latent
    variables and the compactness.
    }
  \label{z_2}
\end{figure*}

\begin{figure*}[]
  \includegraphics[width=\textwidth]{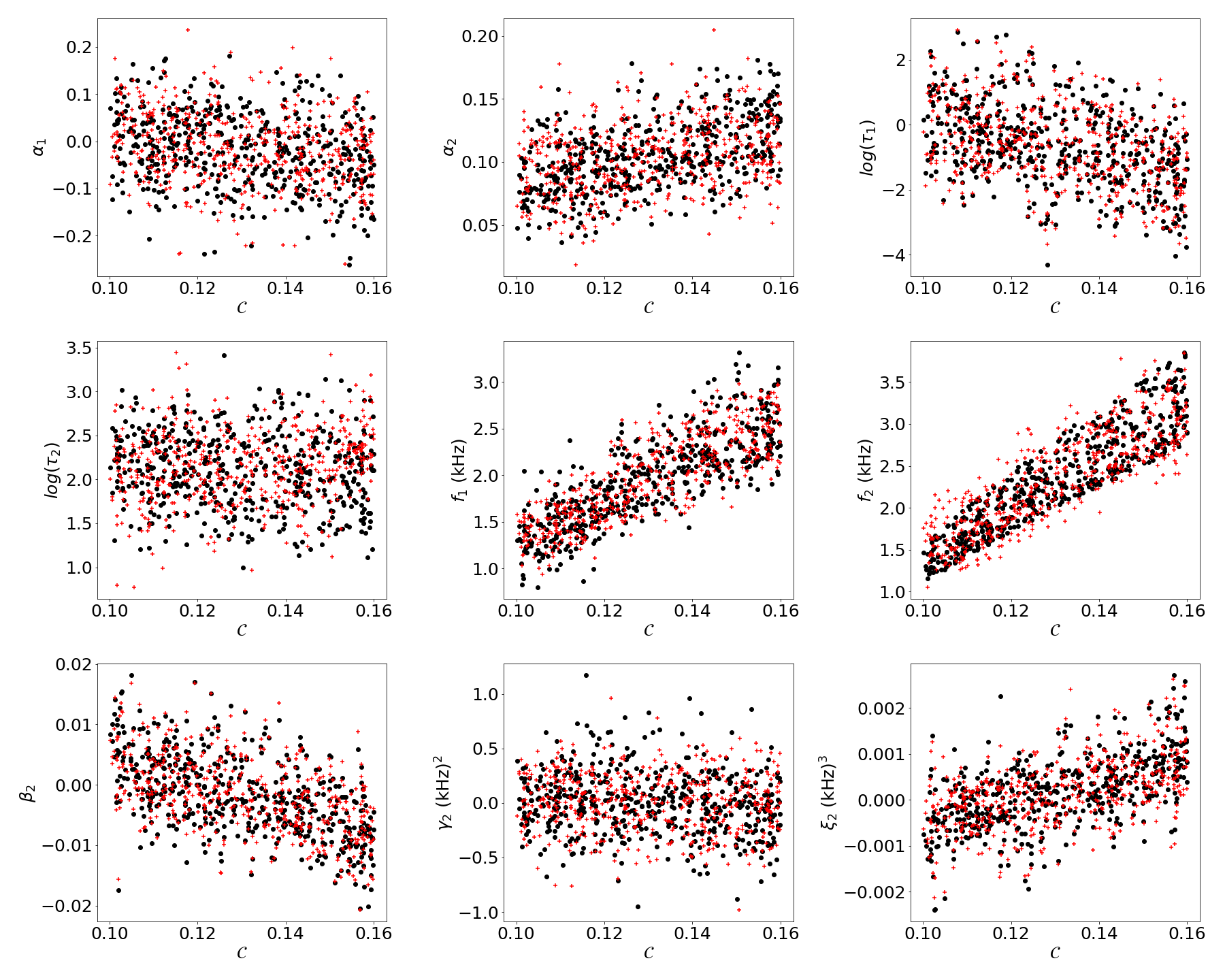}
  \caption{Parameters from the fiducial model with
           $1.2\ M_{\odot} < \bar{M} < 1.7\ M_{\odot} $ and
           $\mathcal{C} \in (0.1,0.16)$ uniformly distributed
	(black dots), compared to the parameters recreated by
	the CVAE (red crosses). The recreated data is 
	generated by inputting the fiducial model 
	data into the CVAE. The CVAE was trained with 
	10,000 elements. Here we fix the 
	value of $\beta = 0.008$, which is different 
	from the $\beta$ value used in Sec.~\ref{sec:Exp}.
    }
  \label{res_param}
\end{figure*}

\begin{figure*}[]
  \includegraphics[width=\textwidth]{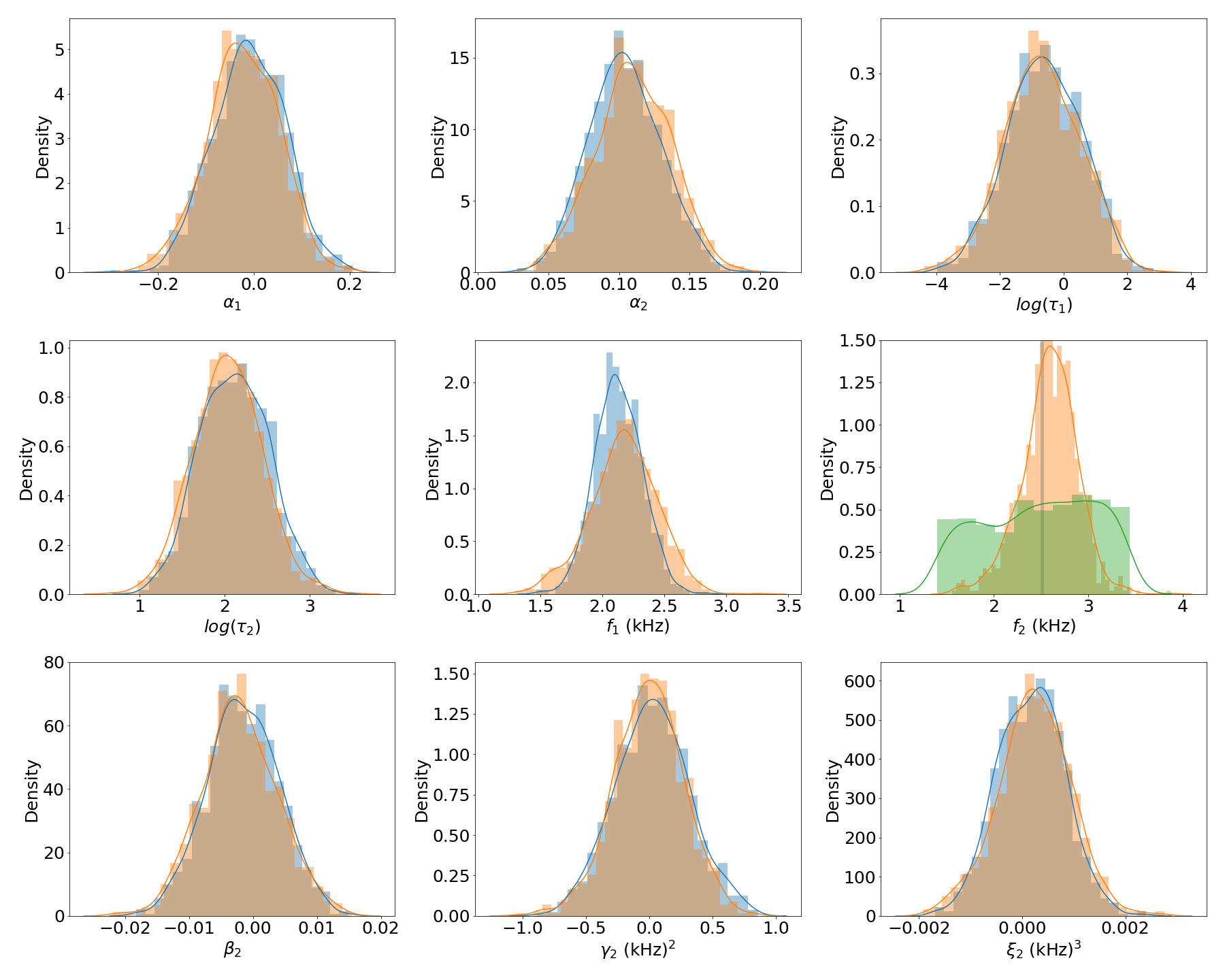}
  \caption{Probability density of fiducial model parameters (blue) and
parameters recreated by the CVAE (orange) with fixed $\mathcal{C} = 0.13$ and
$\bar{M} = 1.35\ M_{\odot}$. For the parameter $f_2$, there is no variance 
in the fiducial model when fixing $\mathcal{C}$ and $\bar{M}$, and for this figure 
we also show the fiducial model distribution with fixed
$\bar{M} = 1.35\ M_{\odot}$, but varying $\mathcal{C}$ (green). The recreated data is
generated by inputting the fiducial model data into the CVAE. The CVAE was
trained with 10,000 elements. Here we fix the value of $\beta = 0.008$, which
is different from the $\beta$ value used in Sec.~\ref{sec:Exp}.
  }
  \label{fixed_c_m}
\end{figure*}

With the fiducial model, we are now able to generate artificial
postmerger signals, which we take as training data for the CVAE. We
first generate a dataset with 1,010,000 samples
$h \sim p_{\text{fid}}(h|\bar M, \mathcal{C})$ by randomly choosing the remnant mass,
$1.2\ M_{\odot} < \bar{M} < 1.7\ M_{\odot} $ and
$\mathcal{C} \in (0.1,0.16)$ from uniform distributions.  The dataset is then whitened and
normalized. We then train CVAEs (with
$l=4$) using 500, 1000, 5000, 8000, 9000, $10^4$, $1.5\times10^4$, $10^5$, and $10^6$ elements. We set aside $10^4$ elements for the validation set. 
We train the CVAE with a batch size of 150 and use early stopping to terminate the training when the loss function becomes stagnant \cite{Goodfellow-et-al-2016}.
Details of the CVAE architecture are provided in Appendix~\ref{paramvalues}. 

%HERE
We find that training the CVAE with a dataset with more than $10,000$ elements
shows close to no signs of overfitting. The match from the input and recreated
waveforms is low, with $\mathcal M \sim 0.27$ for the training set, and $\mathcal M \sim 0.24$
for the validation set. Despite the low matches, we find that the CVAE is still
able to recreate the distribution of the parameters, see Fig.~\ref{res_param},
indicating that the CVAE is functional. As noted above, given the stochastic
nature of the fiducial model, which itself has $\mathcal M \sim 0.38$, 
the match is not the best metric to evaluate the performance of the CVAE.

What is more important then the average match is that the distribution of
parameters reproduced by the CVAE is similar to the ones from the fiducial
model. By looking at Fig.~\ref{res_param}, we see that the recreated parameters
from the CVAE (red crosses) are indeed similar to the input parameters (black dots).
Furthermore, Fig.~\ref{fixed_c_m} shows that the CVAE learns the distributions of the
parameters when fixing the fiducial model arguments
$(\mathcal{C}=0.13,\bar{M}=1.35\ M_{\odot})$. 

The only parameter where the distribution the CVAE recreates differs
significantly from that of the fiducial model is $f_2$, which has no variance
in the fiducial model when fixing $\mathcal{C}$ and $\bar{M}$.  
Though nonzero, the CVAE variance in $f_2$ at fixed $\mathcal{C}$ and $\bar{M}$
is still smaller than the variance of the fiducial model when only fixing $\bar{M}$
(green curve in the $f_2$ panel of Fig.~\ref{fixed_c_m}). 
We also note that when allowing
$\bar{M}$ to vary over the range 1.2 to 1.7 $M_{\odot}$, but still keeping
$\mathcal{C}=0.13$, the standard deviation in $f_2$ for the CVAE is $\sim15\%$ larger than
when $\bar{M}$ is fixed to $1.35\ M_{\odot}$. This is roughly the increase expected
from adding in quadrature the fiducial model variance in $f_2$ when varying $\bar{M}$.

For a given realization of the fiducial model, it is still possible to get
a good match using the CVAE by searching over the latent space.  We found
that, for most cases, a waveform from the CVAE with $\mathcal{M} > 0.90$
could be found by sampling the latent space randomly. And it would
straightforward to implement a more
sophisticated search algorithm for the correct latent variables which
could increase this value. This indicates that the CVAE 
is sufficiently flexible so as to be able to recreate different realizations of the
fiducial model despite the large variance in these realizations representing
large theoretical uncertainties in the the training data.

We can now address the question of how the CVAE encodes information
about the remnant compactness in the latent space. In Fig.~\ref{z_1}, we plot
the latent distribution of encoded waveforms
$h \sim p(\bar M, \mathcal C)$ for fixed
$\bar M = 1.35$ $M_\odot$, but sampling
$\mathcal C \sim p(\mathcal C)$. We see that the latent variables
clearly encode information, as they trace a path in the latent space
rather than sampling the normal distribution. However, there is no
obvious connection to the compactness $\mathcal C$.

To probe whether correlations between $z$ and $\mathcal C$ can be obtained
under more idealized conditions, we train the CVAE once again, but this
time based on a fiducial model with a reduced noise level. Figure~\ref{z_2} shows the latent space of the CVAE trained with the noise
variance reduced by half, i.e., $\Sigma \rightarrow 0.5\Sigma$. We see
that there is a clear correlation between the latent variables 
and $\mathcal C$. The CVAE
has therefore learned about the compactness based on waveforms alone,
without ever having been provided compactness information. This is
evidence of the neural network learning the $\mathcal{C} - f_2$ relationship.
While this is a simple relationship to uncover, it shows how the CVAE can
expose interesting relationships in the latent space.

\section{Discussion}
\label{sec:discuss}

The detection of a gravitational wave signal from the post-merger
oscillations of a BNS would be a major accomplishment with important
consequences for our understanding of physics in extreme regimes.
Given the weaker amplitude and high frequency of such a signal, a good
model will be essential in making and learning from such a detection.
However, our ability to build such a model is hampered by a number of
theoretical uncertainties, including our lack of knowledge of the true
neutron star EOS, uncertainties regarding the behavior of magnetized
matter, viscosity, cooling, and other microphysical effects, and
difficulties adequately resolving turbulence and other small scale
effects, while still covering the allowed parameter space.  Machine
learning offers a way to build partially informed models with
relatively few assumptions about the unknown physics. To build the
CVAE, we assume that the unknown physics can be encoded in a small
latent space, and that there is a way to interpolate between different
theoretical models of the unknown physics.  By training a neural
network with samples covering not only different unknown physics
(e.g. different EOSs), but also different theoretical errors
(e.g. numerical resolutions used in the simulations), we can attempt
to build a model that interpolates over both, and thus is sufficiently
robust to be used to look for real signals.  The neural network
approach also easily adapts to incorporate new information as it is
obtained, and does so with a low associated computational cost at
training time. In this approach, the main cost lies in the amount of
data that is required to train the model. In this case, training the
model with $10^6$ elements took approximately three days on a
laptop. 
This training time can be reduced
significantly by taking advantage of GPUs.

By building CVAEs for toy model data, we showed that the CVAE is
capable of learning simple waveform time series consisting of a sum of sine
functions (see Appendix~\ref{sec:toy_model}). We then moved on to training the CVAE on real numerical waveforms
using the CoRe database \cite{dietrich2018core}, and found that the neural network is
overfitting, which is unsurprising given the dearth of data to train on, and the
complexity of the waveforms.  Instead of training directly on the time series
from the numerical waveforms, we fit a nine parameter function [Eq.~(\ref{rezz})]
to waveforms, and used these parameters as the training data.  We combined this
with some preprocessing to improve the learning outcome, but still found that
we were overfitting. Given that we could not avoid overfitting with the data we
had, we turned our attention to estimating how much data would be required for
the CVAE to work. To perform the estimate, we built a stochastic fiducial model of the
parameters to generate data to train on. 
We generated data
using the fiducial model, trained a CVAE with four latent variables, and found
that we could recreate the distribution of the parameters with $\sim 10,000$ waveforms.
The particular choice of fitting function we used was chosen mainly for
illustration, and other choices, for example, using a greater number of free
parameters, should also work with these methods, though would presumably
increase the required size of the training set.

In investigating the latent space generated by the training process, we found
evidence that the latent space was encoding the hidden variables, but we were
unable to obtain a clear correspondence between the main hidden variable (the
neutron star compactness) and some combination of the latent variables.  A
readily apparent relationship does show up once the noise of the fiducial model is reduced. 
This suggests that in the case without reduced noise, the latent space still
encodes the compactness, but in a more complicated way. This 
relation could perhaps be found by training a secondary neural network for this purpose.  We
leave further study of the latent space to future work.  

In this work, we considered two ways to present the data to the neural
network: the full time series, and the parameters of a model which was
fit to the time series.  However, these are far from the only choices.
Another method which we attempted, but found poor results for, is to
decompose the spectrum of the waveforms with principle component
analysis \cite{Clark:2015zxa}, and then train on the coefficients. We found that we needed
more than 10 principle components to get a match above 0.90. The
combination of large variance and needing many coefficients made the
learning of the data more difficult.

It is also important to keep in mind that the estimate given in
Sec.~\ref{exp} depends heavily on the specific neural network architecture (number of
layers, number of neurons per layer, etc.)~that we train on. The space of
neural network implementations of a CVAE is large, and the one we picked might not be the
optimal one despite our search. One could envision trying to adapt tools such
as autoML \cite{jin2018autokeras}, which searches over the space of neural networks, to
build a CVAE optimal for the post-merger data.

From our studies, we conclude that more training data is required to train the
CVAE to properly represent BNS post-merger waveforms. We estimate that $\sim
10^4$ waveforms are needed. Though one direction for future work would be to
incorporate additional simulation waveforms catalogs
(e.g.~\cite{Kiuchi:2019kzt}), this is still close to two orders of magnitude
more than are presently available.  Further optimizations to the CVAE
architecture could potentially reduce this requirement, as could improved
accuracy in the simulations which would reduce the variance in the waveforms.
We note that here we have attempted to incorporate truncation errors effects in
a simple way by including simulations with the same parameters at multiple
resolution as equal data.  This could be done in a more sophisticated way by
weighting higher resolutions more, or otherwise incorporating truncation error
estimates, though this can be challenging in cases that lack formal convergence
in the discretization scale of the simulation. Given the computational cost
associated with each simulation, accumulating enough waveforms to train a
neural network could be achievable, but would require a large-scale,
coordinated effort from the simulation community.  A properly trained CVAE
could play an important role in mapping out the relationships between different
variables in the space of BNS post-merger waveforms.

\begin{acknowledgments}
T.W., W.E., L.L., and H.Y. acknowledge support from an NSERC Discovery grant. 
L.L. acknowledges CIFAR for support. This research was
supported in part by Perimeter Institute for Theoretical Physics.  Research at
Perimeter Institute is supported by the Government of Canada through the
Department of Innovation, Science and Economic Development Canada and by the
Province of Ontario through the Ministry of Research, Innovation and Science.
This research was enabled in part by support provided by SciNet
(www.scinethpc.ca/) and Compute Canada (www.computecanada.ca).
\end{acknowledgments}

\appendix

 \section{Waveform model with partial information}\label{sec:WMP}
% %
Thanks to numerous developments in solving the equations of general relativity,
the binary black hole waveform is largely known (for comparable mass ratios). 
The corresponding waveform models, whether they are
constructed phenomenologically by matching numerical waveforms with analytical
approximations, or through other methods such as the effective-one-body
approach, are all determined uniquely for every set of binary parameters. On the other hand, to
construct the waveform model for post-merger neutron stars, we are facing a
different problem where our current knowledge is insufficient to predict
an accurate waveform given the orbital parameters.  This problem is likely to
persist until there are significant improvements in our 
understanding of the neutron star EOS (which likely will
come from the detections themselves) and in our modelling of the
merger/post-merger process with a sufficiently complete physical prescription.  

Nevertheless, although current numerical simulations do not provide the full
answer, they do converge on certain features (such as the dominant peak in the
post-merger spectrum).  It is natural to ask how one could construct a waveform
model utilizing this {\it partial} information, while taking into account the
inherent uncertainties. To answer this question, we present a
general framework for waveforms with partial information.

% %
A waveform model with theoretical uncertainties can be written as
$h = h({C};\mathcal{I})$. The control variables ${C}$
are the physical quantities that determine the waveform assuming
perfect knowledge. In the binary case, they are the
orbital parameters, which can be determined from the inspiral waveform
measurement (which in general will be much louder than the post-merger
signal) to a certain accuracy.  The latent variables $\mathcal{I}$
encapsulate all the theoretical uncertainties and modelling errors. In
the limit that a waveform model becomes completely determined without theoretical
uncertainties,
${\rm dim}(\mathcal{I})=0$.

\subsection{Detection}
For the purpose of detection, the control variables are already constrained by
the inspiral measurement. As a result, the posterior distribution of the
control variables $P_{\rm in}({C})$ can be used as a prior distribution
for post-merger detection.  One way to quantify the statistical significance of
a post-merger event detection is using the hypothesis test framework. Using
the post-merger gravitational wave data stream $s$, we compare the following
two hypotheses:
\begin{align}
&\mathcal{H}_1: s =h +n\,,\nonumber \\
&\mathcal{H}_2: s=n\,
\end{align}
where $n$ represents noise.

The significance of detection can be characterized by the Bayes factor, which is defined as
\begin{align}
\mathcal{B} : = \frac{P(s| \mathcal{H}_1)}{P(s| \mathcal{H}_2)}\,,
\end{align}
and the evidence function $P(s| \mathcal{H}_i)$ can be computed by
\begin{align}
P(s| \mathcal{H}_i) = \int d {C} d\mathcal{I} \mathcal{L}_i({C},\mathcal{I}) P_{\rm in}({C}) P_{\rm in}(\mathcal{I})\,,
\end{align}
where $P_{\rm in}(\mathcal{I})$ is the prior weight function for the latent variables. The likelihood function $\mathcal{L}_i$ is given by
\begin{align}
&\mathcal{L}_1  =\frac{1}{Z_n}\prod_{f > 0} {\rm exp}\left ( -\frac{2 |s-h |^2}{S_n}\right )\,,\nonumber \\
&\mathcal{L}_2 = \frac{1}{Z_n} \prod_{f > 0} {\rm exp}\left ( -\frac{2 |s |^2}{S_n}\right ) \,,
\end{align}
where $S_n$ is the single-side spectral density of the detector, and $Z_n$ is a common normalization constant.
The larger the Bayes factor is, the more significant the detection. 
According to the Jeffreys scale of interpretation of Bayes Factor~\cite{jeffries1961theory}, the evidence is ``very strong" if $\mathcal{B}>10$.

Even without prior knowledge of the control
variables, we can still define the SNR of an event as 
\begin{align}
{\rm SNR} : = 2{\rm Max}_{{C},\mathcal{I}} \left (\int df \frac{h({C},\mathcal{I}) s^* + h.c.}{S_n} \right )\,.
\end{align} 
% %
Given a threshold in SNR, it is straightforward to evaluate the average rate of
detecting a false signal due to the detector noise background, which is often
referred to as the false-alarm rate. This part is similar to binary black hole
detection. Generally speaking, for a given false alarm rate, the threshold SNR
is higher if the latent space is larger.

\subsection{Parameter estimation}
% %
While we will not focus on using the model discussed in this paper for
parameter estimation, here we briefly outline how such a 
waveform model with partial information might be used for parameter estimation.  After we compute the
posterior distribution of ${C}$ and $\mathcal{I}$ based on the
gravitational-wave data $s$, the posterior distribution of the control
variables can be obtained by marginalizing the latent variables:
% %
\begin{align}
P({C} | s) = \int d \mathcal{I} P({C}, \mathcal{I} | s)\,.
\end{align}
% %
In reality, the computational cost for constructing the posterior distribution
of ${C}$ and $\mathcal{I}$ increases with the size of the latent space.
Therefore, an efficient waveform model with theoretical uncertainties should minimize the
size of the latent space while still being able to fit the true waveforms.

The latent space encodes information about the unknown physical parameters and
so it is also possible to use the latent space to infer the unknown physical
parameters. This can be accomplished if a map between the unknown physical
parameters and latent space is built. In general this is not an easy task as
the mapping can be quite complicated, but we have seen that this is possible
in Fig.~\ref{z_2} with our fiducial model.

\section{Toy Model}
\label{sec:toy_model}

As a simple demonstration of a CVAE being used as a 
generative model, we train a CVAE on toy data which consists 
of a sum of sine functions.  This is motivated by
the observation that the post-merger waveform seems to be 
dominated by a small number of main frequencies which are 
believed to not vary significantly in time \cite{Clark:2015zxa}.

We generate $\mathcal{O}(10^3)$ waveforms of the form
\begin{equation}
\label{toymodelgen}
w = \sum_{i=1}^3 \sin(\omega_i t + \phi_i)
\end{equation}
where $\phi_i \in [0, 2\pi]$ and $\omega_i \in [1,4]$, $t \in 
[0,15]$ with a sample frequency of $1000/15$. The  phase and 
frequencies are picked randomly from a uniform distribution. 
The data set is then split into a training and a testing set. 
We randomly pick $60\%$ of the total set for training, and 
the other $40\%$ as the testing set. Once the network is trained, we measure the match using Eq.~\ref{match}.

\begin{figure}
\includegraphics[width = 0.5\textwidth]{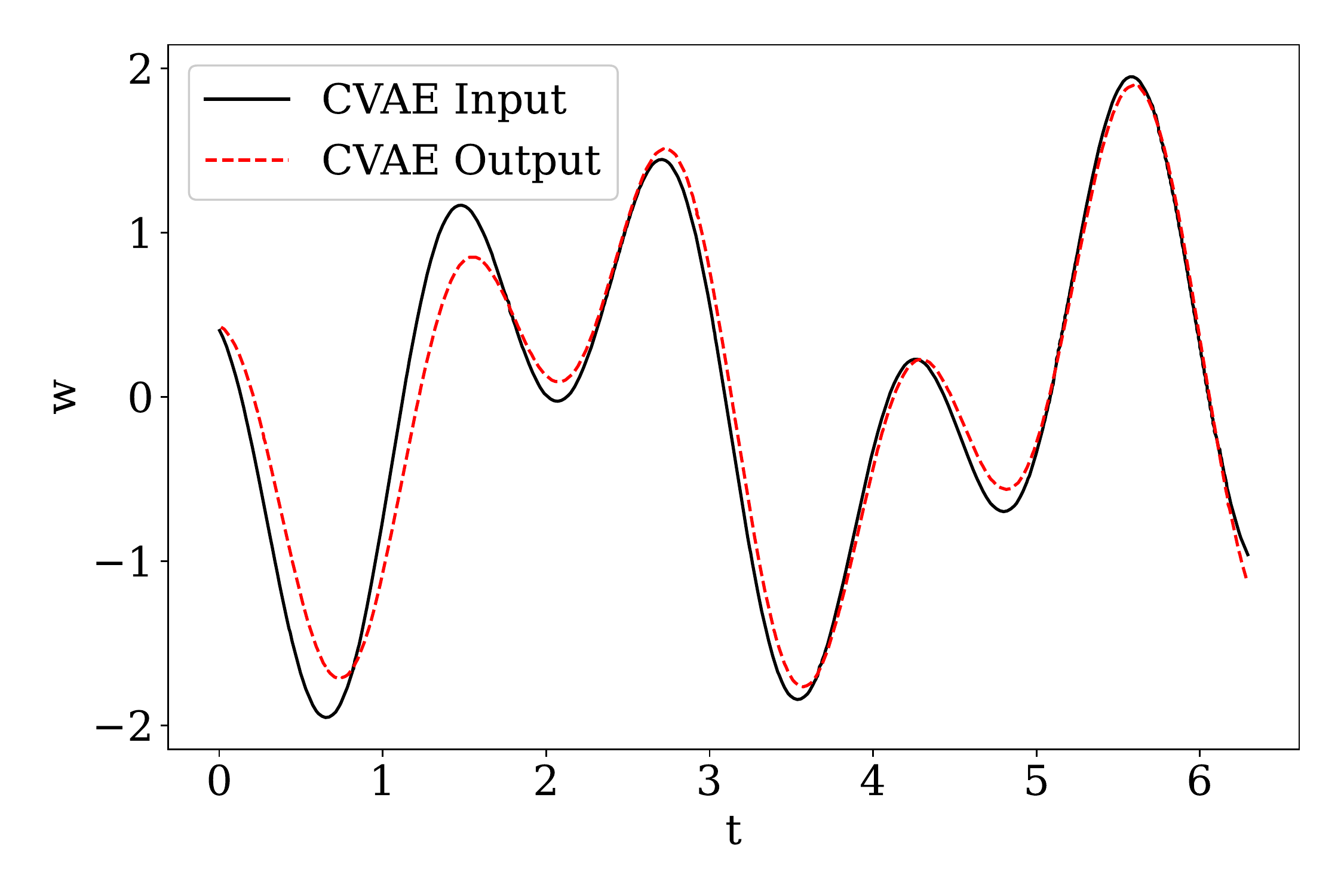}
	\caption{A comparison of the waveform generated by
	Eq.~(\ref{toymodelgen}) (black curve) to the one recreated by the CVAE
	(red curve). The waveform is produced by sampling the latent space and picking the latent variables which produce the highest match. The match is 0.95.
}
\label{TM_095}
\end{figure}

We train the neural network using the frequencies as the 
conditional variables and let the latent space have three 
variables. The neural network has three encoding layers with $\{100,50,10\}$ neurons in the layers. The decoder has the same structure but inverted. The encoder and decoder have rectified linear unit functions as the activation functions, and the latent and output have linear functions as activation functions. Figure \ref{TM_095} is an example output of the 
waveform recreated by the CVAE. We found that the CVAE was 
easily able to recreate the waveform given sufficient amounts 
of data ($\sim 600$ waveforms).

\begin{figure}
\includegraphics[width = 0.5\textwidth]{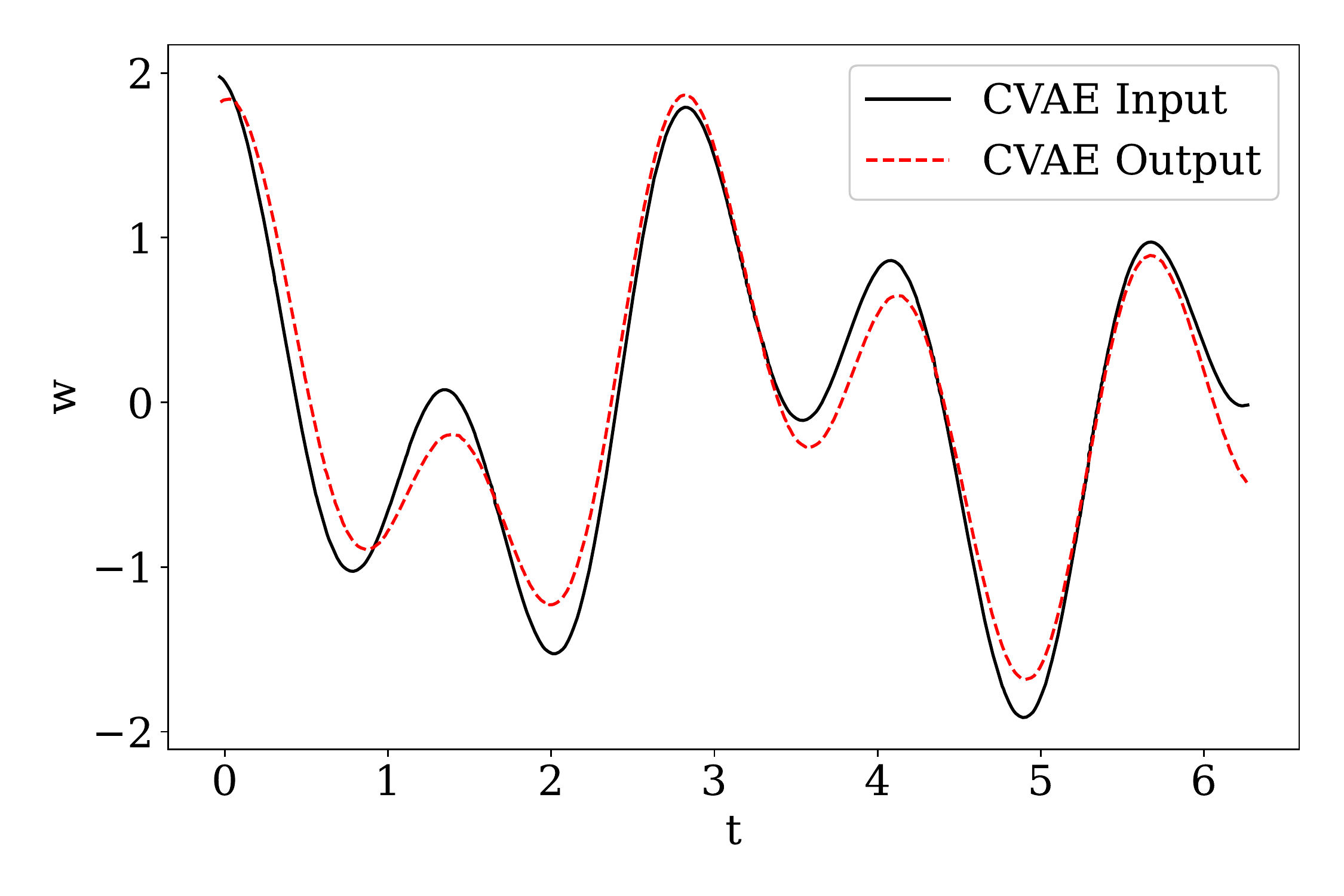}
	\caption{To see that the latent space learned information about the parameters $\phi_i$ we generate a new wave using Eq.~(\ref{toymodelgen}) with the same $\omega_i$ as Fig \ref{TM_095} but different $\phi_i$. We search over our latent space to see if we can also recreate that waveform. Here we see the waveform generated by Eq.~(\ref{toymodelgen}) (black curve) could be recreated by the CVAE (red curve).}
\label{TM_0952}
\end{figure}

\section{Fitting in time domain}\label{FittingTime}

In this section, we discuss the issues of training the CVAE directly on the numerical waveforms. We show that training directly on the numerical waveforms with the amount of data we have leads to overfitting.

We constructed a CVAE similar to the one in Sec.~\ref{toymodelgen}, where the input layer took in the time series of the numerical waveform. The training set consisted of the numerical waveforms and we augmented the dataset by adding random phases to the waveforms. Using this method we increased our dataset to 500 elements.

\begin{figure}[]
\includegraphics[width=0.45\textwidth]{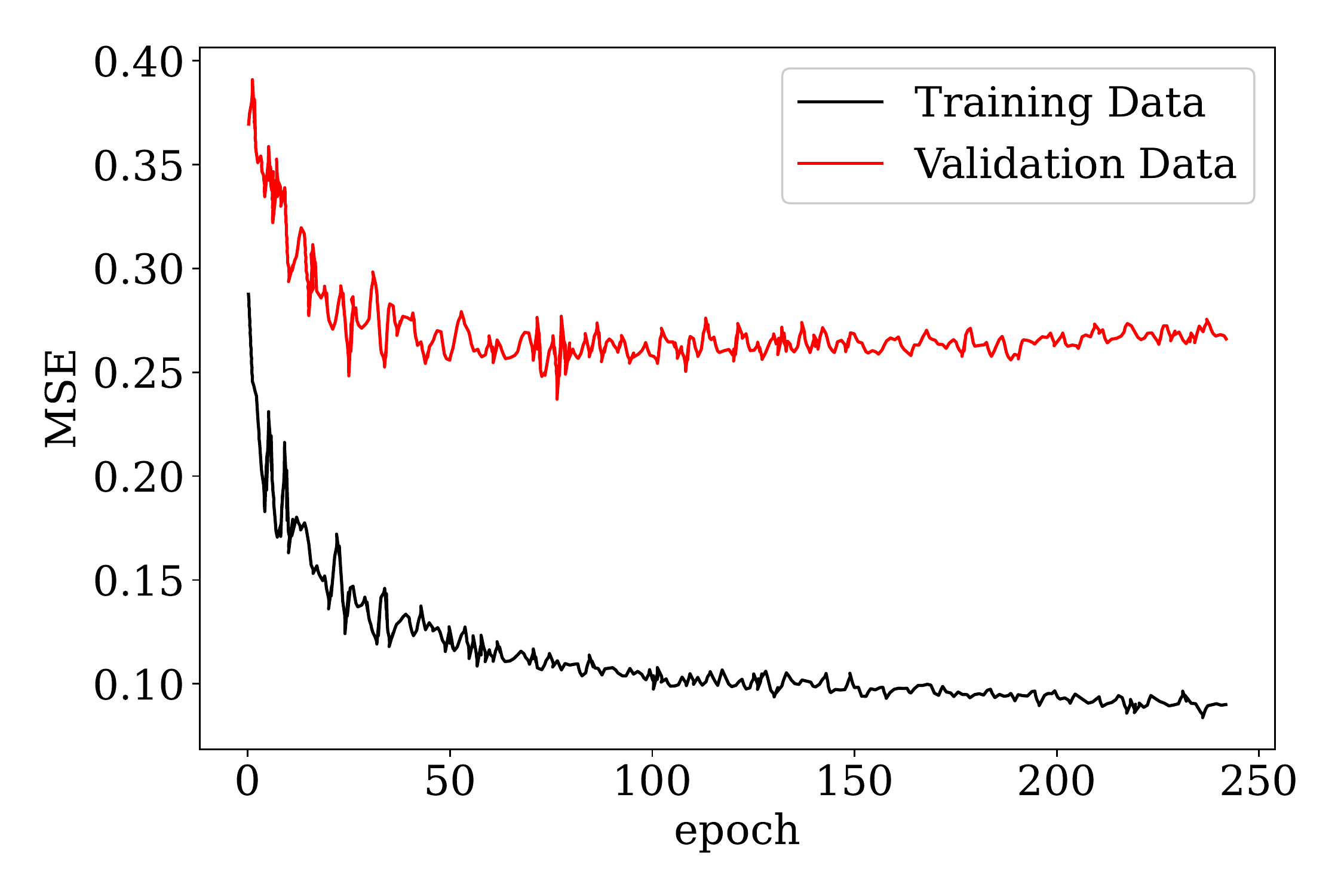}
\caption{
Training history of the CVAE with time domain numerical waveform data. The plot
depicts the mean square error contribution of the loss function, that is how well can it
recreate the waveform.  The difference between the training data and validation
data is a symptom of overfitting.}
\label{train_2}
\end{figure}

After training, we compute the match of the recreated waveforms,
finding that the neural network was in fact 
capable of recreating the waveforms it trained on, but could 
not interpolate between different waveforms. That is, the neural network 
 was overfitting. The overfitting can be seen  
in Fig.~\ref{train_2} where we have the history of 
the mean square error throughout the training. We see that the mean square error of the 
training and testing sets is diverging, indicating that 
the CVAE is learning the training set well, but is not able to 
interpolate.
This is expected given the small amount of training data. In the following
section, we will discuss methods to alleviate this problem, and provide
an estimate of the amount of data required for this model to function. 

\section{Data preprocessing}\label{appendix data process}
In this section, we discuss the details of the data processing done to the data in the main text. The technique used here to preprocess the data is called data whitening; the basic idea is to center the dataset, normalize it, and remove the correlations.

\begin{figure}[H]
	\includegraphics[width=0.5\textwidth]{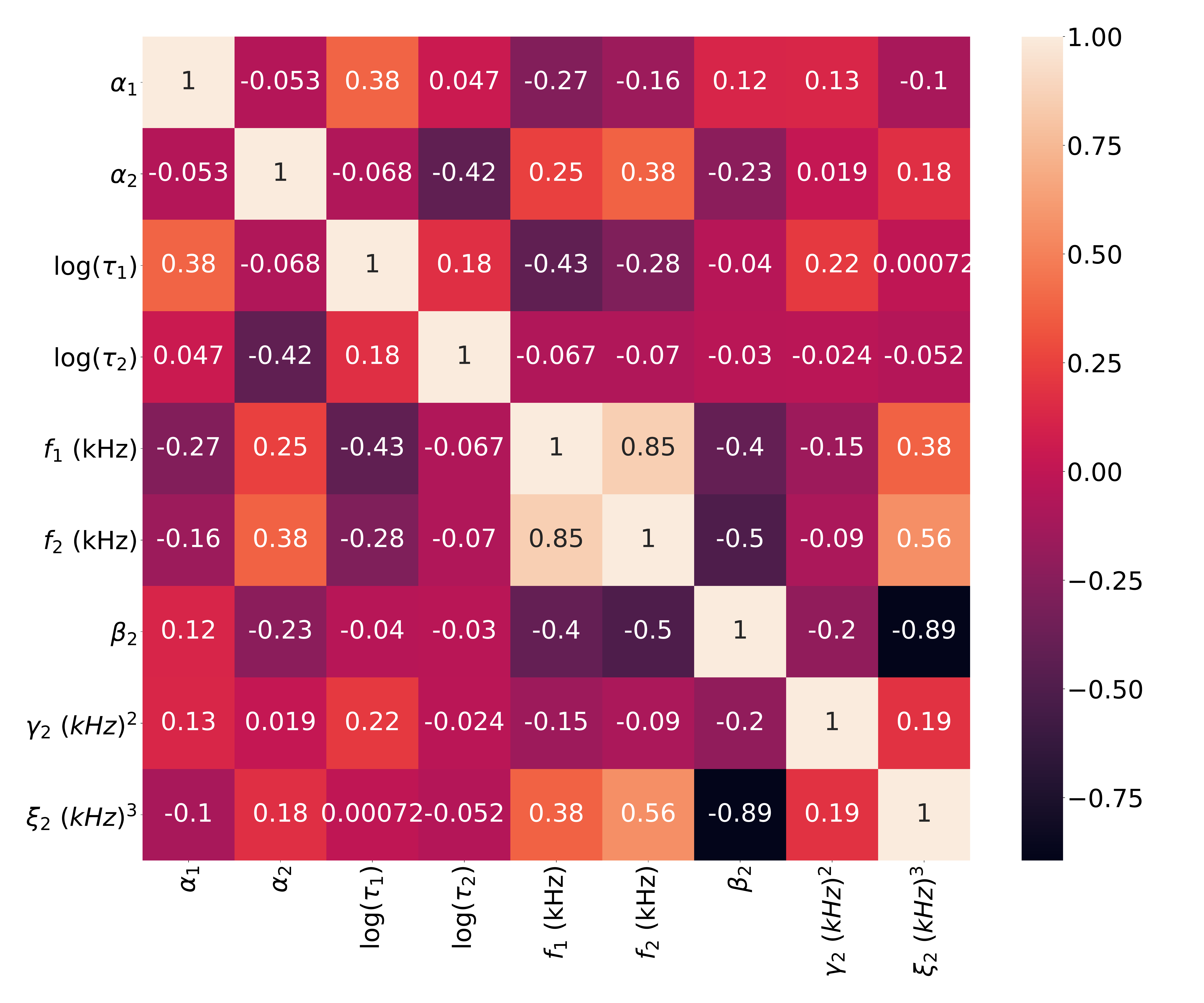}
	\caption{Normalized covariance matrix of the parameters defined in Eq.~(\ref{rezz}),
         when fitted to the numerical relativity waveforms.
	\label{corr}
        }
\end{figure}

Certain subsets of the parameters in
the model given by Eq.~(\ref{rezz}) turn out to be strongly correlated.  This is
illustrated by the correlation matrix (Fig.~\ref{corr}) of the data.
Intuitively, you would suspect that this would make it easier for the CVAE to
reproduce the data. This turns out to not be completely true since it creates
an incentive to only learn the highly correlated parameters and disregard the
uncorrelated ones. This occurs since the correlated terms end up dominating the
mean square term of the cost function. We also found that our data tends toward
posterior collapse when the correlations are left as is. 

To deal with these issues, we can use some basic data processing methods such as
the whitening transform. This transformation consists of three steps: (i)
centering the dataset, (ii) decorrelating the dataset, and (iii) rescaling the
dataset.  To center the dataset, we simply find the mean of every feature and
subtract it from the dataset. This guarantees that the mean of the new dataset
is 0. 
To decorrelate the data we find the eigenvalues $\lambda_i$ and eigenvectors $v_i$ 
of the covariance matrix and define $V = 
[v_1,v_2,...,v_n]$ where the columns are the eigenvectors. The decorrelated data 
$X_{\rm dec}$ is found with respect to the original data 
$X$ by projecting $X$ on $V$:
\begin{equation}
X_{\rm dec} = X \cdot V \ . 
\end{equation}
Finally, to rescale the data, we divide through by the eigenvalues by defining the 
diagonal matrix $P = {\rm diag}(1/\sqrt[]{\lambda_1 + 
\epsilon},1/\sqrt[]{\lambda_2 + \epsilon},...,1/\sqrt[]{\lambda_n + \epsilon})$,
\begin{equation}
X_{\rm whiten} = X_{\rm dec}\cdot P
\end{equation}
where $X_{\rm whiten}$ is the whitened data, and $\epsilon$ is a small number to deal 
with the cases where the eigenvalue is zero (though this does not arise for
the cases we consider). Once the 
data is whitened, we can also tune the parameter $\beta$ in the cost function such that 
no posterior collapse occurs. The tuning of the parameter is done by varying the 
parameter over many training sessions to find the optimal value \cite{BetaVAE}.
After performing these steps, the correlation matrix of the transformed data will just 
be the identity matrix.

\section{Fiducial Model and CVAE Parameters}
\label{paramvalues}
Tables \ref{vecA} and \ref{vecB} give the parameters for Eq.~(\ref{fidModel})
presented in Sec.~\ref{sec:fm} by fitting the numerical waveforms from the CoRe
database \cite{dietrich2018core}. The fits are found by first finding the
compactness using the $f_2$ - $\mathcal{C}$ relation. Then we fit $f_1$ with
the compactness to Eq.~(\ref{eq:f1}), and perform a linear fit to the other
parameters. This gives us the vectors $\vec{A}$ and $\vec{B}$, with the specific
values shown in Tables~\ref{vecA} and~\ref{vecB}. We then compute the covariance
matrix $\Sigma$, finding the values shown in Table~\ref{sigma}. 

\begin{table*}[h]
\center
\begin{tabular}{|l|l|l|l|l|l|l|l|l|l|l|l|l|l|}
\hline
$a_1$ & $a_2$ & $a_3$ & $a_4$ & $a_5$ & $a_6$ & $a_7$ & $a_8$ & $a_9$ & $a_{10}$ & $a_{11}$ & $a_{12}$  & $a_{13}$ & $a_{14}$\\ \hline
-0.99 & 0.78 & 2.73 & 2.41 & 11.26  & -284.83 & 2515.07 & -6800.25 & -3.12 & 51.90 & 89.07 & -0.03 & -2.22 & 0.04\\ \hline
\end{tabular}
\caption{Parameters for $\vec{A}$ in the fiducial model}
\label{vecA}
\end{table*}

\begin{table*}[h]
\center
\begin{tabular}{|l|l|l|l|l|l|l|l|l|}
\hline
$b_1$ & $b_2$ & $b_3$ & $b_4$ & $b_5$ & $b_6$ & $b_7$ & $b_8$ & $b_9$ \\ \hline
0.12 & $3\times 10^{-4}$  & 2.73  & 2.42 & 0.00 & 0.00 & 0.04 & 0.31 & $-5\times 10^{-3}$\\ \hline
\end{tabular}
\caption{Parameters for $\vec{B}$ in the fiducial model}
\label{vecB}
\end{table*}

\begin{table*}[h]
\center
\begin{tabular}{|l|l|l|l|l|l|l|l|l|}
\hline
$8.23\times 10^{-3}$ & $-1.54\times 10^{-4}$ & $4.95\times 10^{-2}$ & $2.19\times 10^{-3}$ & $-4.39 \times 10^{-3}$ & $4.28\times 10^{-3}$ & $1.08\times 10^{-4}$ & $4.85 \times 10^{-3}$ & $-1.20 \times 10^{-5}$ \\ \hline
$-1.54\times 10^{-4}$ & $ 1.00\times 10^{-3}$ &$ -3.04\times 10^{-3}$ & $ -6.87\times 10^{-3}$ & $ -8.23\times 10^{-4}$ &  $-3.41\times 10^{-3}$ & $ -7.05\times 10^{-5}$ & $ 2.55\times 10^{-4}$ & $7.06\times 10^{-6}$ \\ \hline
 $4.95\times 10^{-2} $& $-3.04\times 10^{-3}$ &  $2.01$ &   $1.28\times 10^{-1}$ & $-8.74\times 10^{-2}$ & $1.12\times 10^{-1}$ &  $ -5.41\times 10^{-4}$ & $ 1.29\times 10^{-1}$ &  $1.30\times 10^{-6}$\\ \hline
$2.19\times 10^{-3}$ & $-6.87\times 10^{-3}$ &  $1.28\times 10^{-1} $& $ 2.64\times 10^{-1}$ &$ -3.34\times 10^{-3}$ & $ 8.79\times 10^{-3}$ & $ -1.48\times 10^{-4}$ & $-5.18\times 10^{-3} $& $-3.41\times 10^{-5}$ \\ \hline
$-4.39\times 10^{-3}$ & $-8.23\times 10^{-4}$ & $-8.74\times 10^{-2}$ &
       $ -3.34\times 10^{-3}$ & $2.92\times 10^{-2}$ &$ -2.66\times 10^{-15}$ &
         $5.44\times 10^{-5}$ & $-1.05\times 10^{-2}$ & $-3.34\times 10^{-5}$\\ \hline
$4.28\times 10^{-3}$ & $ -3.41\times 10^{-3}$ &  $1.12\times 10^{-1}$ &
        $ 8.79\times 10^{-3}$ & $-2.66\times 10^{-15}$ & $7.24\times 10^{-2}$ &
         $1.32\times 10^{-3}$ & $ 9.34\times 10^{-3}$ & $-1.96\times 10^{-4}$ \\ \hline
$1.08\times 10^{-4}$ &  $-7.05\times 10^{-5}$ & $-5.41\times 10^{-4}$ &
        $-1.48\times 10^{-4}$ & $ 5.44\times 10^{-5}$ &  $1.32\times 10^{-3}$ &
         $9.19\times 10^{-5} $& $-7.99\times 10^{-4}$ & $-1.09\times 10^{-5}$ \\ \hline
$4.85\times 10^{-3}$ &  $2.55\times 10^{-4}$ &  $1.29\times 10^{-1}$ &
        $-5.18\times 10^{-3}$ & $-1.05\times 10^{-2}$ &  $9.34\times 10^{-3}$ &
        $-7.99\times 10^{-4}$ &  $1.75\times 10^{-1}$ &  $1.01\times 10^{-4}$ \\ \hline
$-1.20\times 10^{-5}$ &  $7.06\times 10^{-6}$ &  $1.30\times 10^{-6}$ &
        $-3.41\times 10^{-5}$ & $-3.34\times 10^{-5}$ &  $-1.96\times 10^{-4} $&
        $-1.09\times 10^{-5}$ &  $1.01\times 10^{-4}$ &  $1.61\times 10^{-6}$ \\ \hline
\end{tabular}
\caption{Elements of the covariance matrix $\Sigma$ in the fiducial model}
\label{sigma}
\end{table*}

Finally, the details of the CVAE used in
Sec.~\ref{exp} are given in Table~\ref{CVAEdetails}.

\begin{table*}[h]
\begin{tabular}{l|l|l|l|l|l|l|l|}
\cline{2-8}
                                          & Input & Layer 1 & Layer 2 & Latent Layer & Layer 3 & Layer 4 & Output \\ \hline
\multicolumn{1}{|l|}{Number of Neurons}   & 9     & 345     & 335     & 4            & 335     & 345     & 9      \\ \hline
\multicolumn{1}{|l|}{Activation Function} & --    & ReLu    & ReLu    & Linear       & ReLu    & ReLu    & tanh   \\ \hline
\multicolumn{1}{|l|}{Type}                &       & Dense   & Dense   & Lambda        & Dense   & Dense   &        \\ \hline
\end{tabular}
\caption{Parameters for the CVAE used in Sec.~\ref{exp}. All the layers in the neural network consist of dense layers, except for the latent space where we have a lambda layer which takes the output of layer 2 as the mean and variance of a multivariate normal and outputs a sample from the multivariate normal. }
\label{CVAEdetails}
\end{table*}

\bibliographystyle{apsrev4-1}
\bibliography{references}

\end{document}